\documentclass[a4paper]{llncs}
\usepackage{graphicx}
\usepackage{cite}
\usepackage{xspace}
\usepackage{color}
\usepackage{url}
\usepackage{subfig}
\usepackage{amsmath}
\usepackage{amssymb}
\usepackage{amsfonts,dsfont}
\usepackage{caption}
\usepackage{verbatim}
\usepackage{multirow}
\usepackage{algpseudocode}
\usepackage{algorithm}
\usepackage{algorithmicx}
\usepackage{adjustbox}
\usepackage{times}
\usepackage[misc,geometry]{ifsym}
\algtext*{EndFor}
\algtext*{EndIf}

\newcommand\PSAN{\textsc{\bf IMP(SAN)}\xspace}
\newcommand\CP{\textsc{\bf CMP}\xspace}

\newcommand {\bQ}{\mbox{\boldmath $Q$}}
\newcommand {\bI}{\mbox{\boldmath $I$}}
\newcommand {\be}{\mbox{\boldmath $e$}}
\newcommand {\E}{\mathcal{E}}

\newcommand {\tr}{ Appendix}

\def\done{\hspace*{\fill} \rule{1.8mm}{2.5mm} }

\urldef{\mailsa}\path|{roczhau,wzylucky}@mail.ustc.edu.cn, {ykli,ylxu}@ustc.edu.cn|
\urldef{\mailsb}\path|hongx87@gmail.com|
\urldef{\mailsc}\path|cslui@cse.cuhk.edu.hk|

\begin{document}

\title{Measuring and  Maximizing Influence via Random Walk in Social Activity Networks}

\titlerunning{}

\author{
\institute{}
}

\author{
Pengpeng Zhao$^1$, Yongkun Li$^{1,2\text{(\Letter)}}$, Hong Xie$^3$, Zhiyong Wu$^1$\\ Yinlong Xu$^{1,4}$, John C. S. Lui$^5$\\
}

\authorrunning{Pengpeng Zhao, Yongkun Li, Hong Xie, Zhiyong Wu, Yinlong Xu, John C. S. Lui.}

\institute{$^1$ University of Science and Technology of China, Hefei, China\\
            \mailsa\\
            $^2$ Collaborative Innovation Center of High Performance Computing, National University of Defense
            Technology, Changsha, China\\
       $^3$ National University of Singapore, Singapore, Singapore\\
\mailsb\\
       $^4$ AnHui Province Key Laboratory of High Performance Computing, Hefei, China\\
       $^5$ The Chinese University of Hong Kong, Hong Kong, China\\
       \mailsc\\
       }

\maketitle

\begin{abstract}
With the popularity of OSNs, finding a set of most influential users (or nodes) so
as to trigger the largest influence cascade is of significance.
For example, companies may take advantage of the ``word-of-mouth'' effect to
trigger a large cascade of purchases by offering free samples/discounts to
those most influential users. This task is usually modeled as an influence maximization problem,
and it has been widely studied in the past
decade. However, considering that users in OSNs may participate in various
kinds of online activities, e.g., giving ratings to products, joining discussion
groups, etc.,   influence diffusion  through online activities becomes even
more significant.

In this paper, we study the impact of online activities by formulating the
influence maximization problem for social-activity networks (SANs)
containing both users and online activities. To address the computation
challenge, we define an influence centrality via random walks to measure
influence,  then  use the Monte Carlo framework to efficiently estimate
the centrality  in SANs. Furthermore, we develop a greedy-based algorithm
with two novel optimization techniques to find the most influential users.
By conducting extensive experiments with real-world datasets, we show our
approach is  more efficient than the state-of-the-art algorithm IMM
\cite{tang2015influence} when we needs to handle large amount of online
activities.


\end{abstract}

\section{Introduction}\label{sec:introduction}

Due to the popularity of online social networks (OSNs), viral marketing
which exploits the ``word-of-mouth'' effect is of  significance to
companies which want to promote product sales.
Therefore, it is  of interest  to find the best initial set of users so as to trigger the
largest influence spread.
This viral marking problem can be modeled as an influence maximization problem, which was first formulated by Kempe et al. \cite{kempe2003maximizing}.
That is, given an OSN and an information diffusion model, how to select a set of $k$
users, which is  called
 the seed set, so as to trigger the largest influence spread. This problem is
 proved to be an NP-hard problem \cite{chen2010scalable, chen2010scalableICDM},
and it has been studied extensively in the
 past decade \cite{chen2009efficient, chen2010scalable, chen2010scalableICDM,tang2014influence, tang2015influence}.

\begin{figure}[!t]
\centering\includegraphics[width=0.5\linewidth]{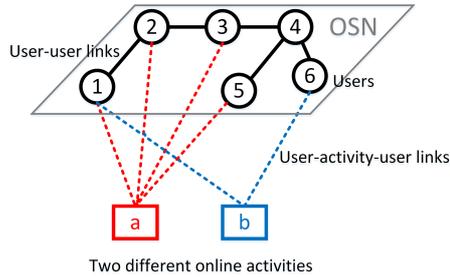}
\vskip -10pt
\caption{An example of social-activity network (SAN).}
\vskip -20pt
\label{fig:SAN}
\end{figure}

Note that users in today's OSNs may participate in various kinds of
online activities, e.g.,
joining a discussion group, and clicking \texttt{like} on Facebook etc.
Hence, users  not only can create friendship relationships, which we call {\em
user-user} links, but can also form relationships by  participating in
online activities, which we call {\em user-activity-user} links. For
example, if two users in Facebook express like to the same public page, then
they form a user-activity-user link no matter they are friends or not. We
call this kind of networks which contain both user-user relationships and
user-activity-user relationships as {\em social-activity networks} (SANs).

With the consideration of online activities in SANs,
influence may also spread through the user-activity-user
links as well as the user-user links.
In this paper, we focus on the
online activities which generate positive influence, e.g., clicking
\texttt{like} on the same public page in Facebook, giving high rating to the
same product in online rating systems, and joining in a community sharing
the same interest in online social networks.
Due to the large amount of online activities, e.g., each pair of users may
participate in multiple online activities,  influence diffusion through the
user-activity-user links becomes even more significant, and so
only considering OSNs alone may not trigger the largest influence spread.
Existing works  on influence maximization  usually focus on OSNs only and do
not take the impact of online activities into consideration. {\em This
motivates us to formulate the influence maximization problem for SANs,
and  to determine the most influential nodes by  taking online activities into
consideration.}

However, solving the influence maximization problem in SANs with online activities
 is  challenging. First, influence maximization in
OSNs without online activities was already proved to be NP-hard, and  considering
online activities
 makes this problem even more complicated. Second, the amount of online activities
 in a SAN is  very large even for small OSNs, this is
 because  online activities  happen  more frequently than
 friendship formation in OSNs. As a result, the underlying graph which characterizes users  and their relationships  may become
 extremely dense if we transform the user-activity-user links to user-user links,
 so it requires  highly efficient algorithms for finding the most influential
 nodes.
To address the above challenges, in this paper,
we make the following contributions.

\begin{itemize}
    \item{We generalize the influence diffusion models for SANs by
        modeling SANs as hypergraphs, and  approximate the influence of
        nodes in SANs by defining an influence centrality based on random
        walk.}
    \item{We  employ  the Monte Carlo framework to  estimate the influence
        centrality in SANs, and also develop a greedy-based algorithm with
        two novel optimization techniques to solve the influence
        maximization problem for SANs. }
    \item{  We conduct  experiments with real-world datasets, and results
        show that our approach is more efficient while keep almost the
        same accuracy compared to the
      state-of-the-art algorithm. }
\end{itemize}

This paper is organized as follows. In \S\ref{sec:formulation}, we formulate
the influence maximization problem for SANs. In \S\ref{sec:methodology}, we
present our random walk based methodology. In \S\ref{sec:computation}, we
present the Monte Carlo method to estimate the influence centrality in SANs.
In \S\ref{sec:maximization}, we present our greedy-based algorithm and
optimization techniques to solve the influence maximization problem. In
\S\ref{sec:experiment}, we present the experimental results. Related work is
given in \S\ref{sec:related} and \S\ref{sec:conclusion} concludes.

\section{Problem Formulation}\label{sec:formulation}

In this section, we first model the SAN with a hypergraph, and then
formulate the influence maximization problem for SANs.

\vspace{-5pt}
\subsection{Model for SANs}\label{subsec:san}

We use a hypergraph $G(V, E, \E_1,...,\E_{l})$ to characterize a SAN, where
$V$ denotes the set of users, $E$ denotes the {\em user-user} links, and
$\E_i$ ($i=1,2,...,l$) denotes the set of type $i$ hyperedges in which each
hyperedge is a set of users who participated in the same online activity,
and represented as a tuple. Considering Figure~\ref{fig:SAN}, only activity
$a$ is of the first type, so $\E_1=\{(1,2,3,5)\}$. For ease of presentation,
we denote $N(j)$ as the set of neighbors of user $j$, i.e.,
$N(j)=\{i|(i,j)\in E\}$, $M_e(j)$ as the set of users except for user $j$
who connected to the hyperedge $e$, i.e., $M_e(j)=\{i | i \in e, \& i\neq
j\}$, and denote $\mathcal{E}_t(j)$ as the set of type $t$ hyperedges that
are connected to user $j$, i.e., $\mathcal{E}_t(j)=\{e|e\in \E_t \& j\in e
\}$. Considering Figure~\ref{fig:SAN}, $N(1)=\{2\}$, $M_e(1)=\{2,3,5\}$ when
$e=(1,2,3,5)$ and $\mathcal{E}_1(1)=\{(1,2,3,5)\}$.

\vspace{-5pt}

\subsection{Influence Maximization in SANs}\label{subsec:im_san}

Before describe the influence diffusion process for SANs,
we first recall the independent cascade model(IC) which was proposed by
Kempe et al. in \cite{kempe2003maximizing}.
Suppose that each user has two states, either active or inactive.
At first, we initialize a set of users as active. For an active user $i$, she will activate each of
her inactive neighbor $j$ ($j\in N(i)$ where $N(i)$ denotes the neighbor set
of user $i$) with probability $q_{ij}$ ($0\leq q_{ij}\leq 1$).
 One common setting of $q_{ij}$ is $q_{ij}=\frac{1}{d_j}$, e.g.,
\cite{kempe2003maximizing, chen2009efficient, chen2010scalable,
tang2014influence, tang2015influence}, where $d_j$ denotes the degree of
user $j$, i.e., $d_j=|N(j)|$. After a neighbor $j$ being activated, then she
will further activate her inactive neighbors in the set $N(j)$, and this
diffusion process continues until no user can change her state.
We call the expected size of the final set of active users the {\em
influence spread}, and denote it as $\sigma(S(k))$ if the set
 of $k$ initial active users is $S(k)$.

Now we describe the influence diffusion process for SANs. The
key issue is to define the influence between user $i$
and user $j$ (i.e., $g_{ij}$) after taking online activities into consideration. Our
definition is based on three criteria:

\begin{itemize}
  \item A user  may make a purchase due to her own interest or being
      influenced by others through  user-user  or user-activity-user
      links, so we define the total influence probability by one-hop
      neighbors as $c$ ($0<c<1$), and call it the decay parameter. As we
      have $l$ types of online activities,
we define $\alpha_{jt}$
       (where $0\leq \alpha_{jt} \leq 1$ and
  $0\leq
      \sum_{t=1}^l\alpha_{jt} \leq 1$) as the proportion of influence to
       user $j$  through  type $t$ online activities, and call it {\em
       weight of activities}. Clearly,
$1-\sum_{t=1}^l\alpha_{jt}$ indicates the proportion of influence from
direct neighbors.

\item For the influence to user $j$ from direct neighbors, we define the
    weight of each neighbor $i$ ($i\in N(j)$) as $u_{ij}$, and assume that
    $0\!\leq\! u_{ij}\!\leq \!1$ and  $\sum_{i\in N(j)}u_{ij}\!=\!1$.

\item   For the influence to user $j$  through the type $t$ online
    activities, we define the weight of each online activity $a$ as
    $v_{aj}$, where $0\leq v_{aj}\leq 1$ and $\sum_{a\in N_t(j)}v_{aj}=1$.
    Besides, considering that maybe multiple users participated in the
    same online activity $a$,  we define the weight of each user $i$ who
    participated in $a$ as  $u^a_{ij}$ ($i \in
                       N(a)\backslash\{j\}$), and  assume that $0\leq
                       u^a_{ij}\leq 1$ and $\sum_{i \in
                       N(a)\backslash\{j\} }u^a_{ij}=1$.

\end{itemize}

\vskip -5pt For simplicity, we let $u_{ij}= 1/|N(j)|$ in this paper. Note
that this uniform setting is
    exactly the same as the IC model in OSNs, which has been widely studied in
    \cite{kempe2003maximizing, chen2009efficient, chen2010scalable,
    tang2014influence,
tang2015influence}.  Similarly, we also let $v_{aj}=1/|\mathcal{E}_t(j)|$
and $u^a_{ij}=1/|M_e(j)|$ by following the uniform setting. We would like to
point out that our random walk approach in this paper also applies to
general settings. Now we can define the influence of user $i$ to user $j$,
which we denote as $g_{ij}$: \vspace{ -5pt}
\begin{equation}
\small
   g_{ij}\! =\! c \times \!\Big(\! \frac{1-\sum_{t=1}^l\alpha_{jt} } {|N(j)|} \times \textbf{1}_{\{i\in N(j)\} } +\!\!
     \sum\limits_{t\in[1,l]}  \! \sum\limits_{e\in \mathcal{E}_t(j) } \!\!\frac{\alpha_{jt}}{|\mathcal{E}_t(j)|}
     \times  \frac{1}{|M_e(j)| }\!\times\! \textbf{1}_{\{i\in M_e(j)\} } \Big)  .
   \label{eq:general_im}
\end{equation}
The first part in the right hand side of Equation~(\ref{eq:general_im})
denotes the influence diffusion through user-user links, and the second part
represents the influence diffusion through user-activity-user links.

Now we formulate the influence maximization problem for SANs, which we denote as \PSAN,
 as follows.
\begin{definition}
\noindent\PSAN: Given a SAN $G(V, E, \E_1,...,\E_{l})$, an influence
diffusion model  with parameters $\alpha_{jt}$, find a set of $k$ nodes
$S(k)$, where $k$ is an integer, so as to make the influence spread
$\sigma(S(k))$ maximized.
\end{definition}
\vskip -15pt

\section{Methodology}\label{sec:methodology}

In this section,  we present our methodology to address the  (\PSAN)
problem. To reduce the large computation cost,  we first develop a random
walk framework on  hypergraphs to estimate the influence diffusion process.
Then, we define a centrality measure based on random walk
 to approximate the influence of a node set. With this centrality measure, we can approximate the influence
maximization problem by solving a centrality maximization problem.

\vspace{-5pt}
\subsection{Random Walk on Hypergraph} \label{subsec:rw_hypergraph}

Here, we present our random walk based framework, which is
extended from the classical
random walk on a simple unweighted graph $G(V,E)$, which can be stated as
follows. For a random walk at vertex $i \in V$, it uniformly selects at random
 a neighbor $j$ ( $j\in N(i)$),  and then moves to
 $j$ in the next step. Mathematically, if we denote $Y(t)$ as the
position of the walker at step $t$, then $\{Y(t)\}$ constitutes a Markov
chain where the one-step transition probability $p_{ij}$ is defined as
$p_{ij}= 1/|N(i)|$ if $(i,j) \in E$, and 0 otherwise.

We now define the one-step transition probability $p_{ij}$ when performing a
 random walk on the hypergraph $G(V,E,\E_1,...,\E_{l})$. Note that each
 hyperedge may contain more than two vertices, so we take the
 one-step random walk from user $i$ to user $j$ as a two-step process.

\noindent
$\bullet$
{\bf Step one:} Choose a hyperedge associated to user $i$. Precisely, according to the influence diffusion models in \S\ref{subsec:im_san},
we  set the probability of
       selecting  type $t$ hyperedges as $\alpha_{it}$,  and choose
        hyperedges of the same type uniformly at random. Mathematically, if
       the walker is currently at user  $i$, then it chooses a hyperedge $e$ of
       type $t$ with probability $\frac{\alpha_{it}}{|\mathcal{E}_t(i)|}$.

\noindent
$\bullet$ {\bf Step two:} Choose a user associated to the hyperedge $e$ selected in
       step one as the next stop of the random walk. We consider random walks without backtrace. In particular, if a walker is currently at node $i$,
       then we  select the next stop uniformly from the vertices that are connected to the same hyperedge with user $i$.
       We define the probability of choosing
       user $j$ as $1/|M_e(i)|$.

 By combing the two steps defined above,  we can derive the transition probability from user $i$ to $j$ as follows, and we can find
  $g_{ji} = c\times p_{ij}$.
\begin{footnotesize}
  \begin{align}
   p_{ij} &= \frac{1-\sum_{t=1}^l\alpha_{it} } {|N(i)|} \times \textbf{1}_{\{j\in N(i)\} } +
     \sum\limits_{t\in[1,l]}   \sum\limits_{e\in \mathcal{E}_t(i) } \frac{\alpha_{it}}{|\mathcal{E}_t(i)|}   \times  \frac{1}{|M_e(i)| }\times \textbf{1}_{\{j\in M_e(i)\} }  .
   \label{eq:transition_prob}
 \end{align}
\end{footnotesize}

\vspace{-15pt}

\subsection{Influence Centrality Measure}\label{subsec:centrality_def}

To address the  \PSAN problem, one key issue is to measure the influence of
 a node set. To achieve this,  we define a centrality measure based on  random
walks  on hypergraphs to approximate the influence of a node set $S$. We
call it {\em influence centrality}, and denote it as $I(S)$, which is
defined as follows.
\begin{equation}
I(S)=\sum\nolimits_{j\in{V}}h(j,S),
\label{eq:influence_centrality}
\end{equation}
where $h(j,S)$ aims to approximate the influence  of $S$ to  $j$, which is
called decayed hitting probability. It is defined as
\begin{equation}
h(j,S)=
\begin{cases}
\sum_{i \in V}cp_{ji}h(i,S), j \notin S, \\
1, j \in S,
\end{cases}
\label{eq:decayed_hitting_prob}
\end{equation}
where $c$ is the decay parameter defined in \S\ref{subsec:im_san},
and $p_{ji}$ is the one-step transition probability  defined in Equation~(\ref{eq:transition_prob}).

To solve the influence maximization problem of \PSAN, we  use the influence
centrality measure $I(S)$ to approximate the influence of the node set $S$,
and our goal is to find a set $S$ of $k$ users so that $I(S)$ is maximized.
In other words, we approximate the influence maximization problem \PSAN by
solving the centrality maximization problem  \CP
 defined as follows.

\begin{definition}
\noindent \CP: Given a hyperghraph $G(V,E,\E_1,...,\E_{l})$ and the
corresponding parameters $\alpha_{jt}$, find a set $S$ of $k$ nodes, where
$k$ is an integer, so as to make the influence centrality of the set $S$ of
$k$ nodes $I(S)$ maximized.
\end{definition}

\section{Centrality Computation}\label{sec:computation}
We note that the key challenge of solving the centrality maximization problem
\CP is how to efficiently estimate the influence centrality of a node set
$I(S$), or  the decayed hitting probability $h(j,S)$.
We give an efficient framework to estimate $h(j,S)$ as follows.
We first rewrite $h(j,S)$ in  a  linear  expression
which is an infinite converging series, and then truncate the converging
series to save computation time (see \S\ref{subsec:recursive}). To further
estimate the truncated series, we first explain the expression with a random
walk approach, and then use a Monte Carlo framework via random walks to
estimate it efficiently (see \S\ref{subsec:monte_carlo}).

\subsection{Linear Expression}\label{subsec:recursive}

We first transform $h(j,S)$ defined in
Equation~(\ref{eq:decayed_hitting_prob}) to a linear  expression.

\begin{theorem}
The decayed hitting probability $h(j,S)$   can be rewritten as
\begin{equation}
h(j,S)=c\be_{j}^T\bQ'\be+c^2\be_{j}^T\bQ\bQ'\be+c^3\be_{j}^T\bQ^2\bQ'\be+\cdots.
\label{eq:infinit_series}
\end{equation}
where $\bQ$ is a $(|V|-|S|)\times (|V|-|S|)$ dimensional matrix which
describes the transition probabilities between two nodes in  the set $V-S$,
$\bQ'$ is a $(|V|-|S|)\times |S|$ dimensional  matrix which describes the
transition probabilities from a node in $V-S$ to a node in $S$, $\bI$ is an
identity matrix, $\be$ is a column vector with all elements being 1, and
finally $\be_j$ is a column vector with only the  element corresponding to
node $j$ being 1 and 0 for all other elements.
\end{theorem}

{\noindent \bf Proof: } Please refer to the\tr. \done

We only keep the $L$ leading terms of the infinite series, and
denote the truncated result as $h^L(j,S)$, so we have
\begin{equation}
h^L(j,S)=c\be_{j}^T\bQ'\be+c^2\be_{j}^T\bQ\bQ'\be+...+c^L\be_{j}^T\bQ^{L-1}\bQ'\be.
\label{eq:truncated_series}
\end{equation} 
Since $c$ is defined as $0<c<1$, the series truncation error is bounded as
follows.
\begin{equation}
0 \leq h(j,S)-h^{L}(j,S) \leq c^{L+1} / (1-c).
\label{eq:truncation_error_bound}
\end{equation}

Based on the above error bound,  we can see that $h^L(j,S)$ converges to
$h(j,S)$ with rate $c^{L+1}$. This implies that if we want to compute
$h(j,S)$ with a maximum error $\epsilon$ ( $0\leq \epsilon \leq 1$), we only
need to compute $h^L(j,S)$ by taking a sufficiently large enough $L$, or $L\geq \lceil
\frac{\log (\epsilon-\epsilon c)}{\log c} \rceil-1$.

\subsection{Monte Carlo Algorithm}\label{subsec:monte_carlo}

In this subsection, we present  a Monte Carlo  algorithm to efficiently
approximate $h^{L}(j,S)$. Our algorithm is inspired from the random walk
interpretation of Equation~(\ref{eq:truncated_series}), and it can achieve a
high accuracy with  a small number of walks.

Consider the random walk interpretation of a particular term
$\be_{j}^T\bQ^{t-1}\bQ'\be$ ($t=1,...,L$) in
Equation~(\ref{eq:truncated_series}). Let us consider a $L$-step random walk
starting from $j \notin S$ on the hypergraph. At each step, if the walker is
currently at node $k$ ($k \notin S$), then it selects a node $i$ and
transits to $i$ with probability $p_{ki}$, which is defined in
Equation~(\ref{eq:transition_prob}). As long as the walker hits a node in
$S$, then it stops. Let $j^{(t)}$ be the $t$-th step position, and define an
indicator $X(t)$ as
\begin{equation*}
X(t)=
\begin{cases}
1 , \quad j^{(t)} \in S, \\
0 , \quad j^{(t)} \notin S.
\end{cases}
\end{equation*}

We can see that  $\be_{j}^T\bQ^{t-1}\bQ'\be$ is the probability that a
random walk starting from $j$ hits a node in $S$ at the $t$-th step.
We have
\begin{equation}
\be_{j}^T\bQ^{t-1}\bQ'\be=E[X(t)].
\label{eq:rw_interpretation}
\end{equation}
\noindent By substituting $\be_{j}^T\bQ^{t-1}\bQ'\be$ with
Equation~(\ref{eq:rw_interpretation}), we can rewrite $h^L(j,S)$ as
\begin{equation}
h^L(j,S)=cE[X(1)]+c^2E[X(2)]+\cdots +c^LE[X(L)].
\label{eq:rw_sum}
\end{equation}

Now we estimate $h^L(j,S)$ by using a Monte Carlo method with random
walks on the hypergraph based on Equation~(\ref{eq:rw_sum}).  Specifically,
for each node $j$ where $j\notin S$, we set $R$ independent $L$-step random
walks  starting  from $j$.
We denote the $t$-th step position of the $R$
random walks  as $j_1^{(t)}$, $j_2^{(t)}$, ... ,
$j_R^{(t)}$, respectively, and use $X_r{(t)}$ to indicate whether
$j_r^{(t)}$ belongs to set $S$ or not. Precisely, we set $X_r{(t)}= 1$ if
$j_r^{(t)}\in S$, and 0 otherwise,
so  $c^tE[X(t)]$  can be estimated as
\begin{equation*}
c^tE[X(t)] \approx \frac{c^t}{R}\sum\nolimits_{r=1}^{R}X_r{(t)}.
\end{equation*}
By substituting $c^tE[X(t)]$  in Equation~(\ref{eq:rw_sum}),  we can
approximate $h^L(j,S)$, which we denote as $\hat h^L(j,S)$, as follows.
\begin{equation}
\hat h^L(j,S) = \frac{c}{R} \sum\nolimits_{r=1}^{R}\!X_r{(1)} +\cdots+
\frac{c^L}{R} \sum\nolimits_{r=1}^{R} X_r{(L)}.
\label{eq:rw_approx}
\end{equation}

Algorithm~\ref{alg:alg1} presents the process of the Monte Carlo method
described above. We can  see that its time complexity is $O(RL)$ as the
number of types of online activities $l$ is usually a small number. In other
words, we can estimate $h^L(j,S)$ in $O(RL)$ time and compute $I(S)$ in
$O(nRL)$ time as we need to estimate $h^L(j,S)$ for all nodes. The main
benefit of this Monte Carlo  algorithm is that its running time is
independent of the graph size, so it scales  well to large graphs.

\begin{algorithm}[!htb]
    \caption{Monte Carlo Estimation for $h^L(j,S)$}
    \small
    \begin{algorithmic}[1]
	 \Function{$h^L(j,S)$}{}
	 	\State $\sigma \leftarrow 0$;
	 	\For{ $r$ = 1 to $R$}
            \State $i \leftarrow j$;
	 		\For{ $t$ = 1 to $L$}
                \State Generate a random number $x \in [0,1]$;
                \For{$T$ = 0 to $l$}
                    \If{$x \leq \alpha_{iT}$}  \Comment{$\alpha_{0T} =
                    1-\sum_{T=1}^l\alpha_{iT}$;}
                        \State $E \leftarrow \mathcal{E}_T(i)$;
                        \State break;
                    \EndIf
                    \State $x \leftarrow x-\alpha_{iT}$;
                \EndFor
                \State Select a hyperedge $e$ from $E$ randomly;
                \State $i\! \leftarrow\!$ select a user from $\{k | k\!\!\in\! \!e, k\!\neq\!
                i\}$ randomly;
                \If{$i \in S$}
 				    \State $\sigma \leftarrow \sigma + c^{t}/R$;
                    \State break;
                \EndIf
			\EndFor
		\EndFor
		\State \Return{$\sigma$};
	 \EndFunction
\end{algorithmic}
 \label{alg:alg1}
\end{algorithm}

 Note that $\hat h^L(j,S)$
computed with Algorithm~\ref{alg:alg1} is  an approximation of
 $h^L(j,S)$, and the approximation error depends on the sample size $R$.
To estimate the number of samples required to compute $h^L(j,S)$ accurately,
we derive the error bound by applying Hoeffding inequality
\cite{hoeffding1963probability}, and the results are as follows.

\begin{theorem}
Let the output of Algorithm~\ref{alg:alg1} be $\hat h^L(j,S)$, then we have
\begin{equation}
P\{|\hat h^L(j,S)-h^L(j,S)| \!>\! \epsilon \} \!\leq\! 2L\exp(-2 (1-c)^2{\epsilon}^2R).
\end{equation}
\vspace{-0.2in} \label{theo:error_bound}
\end{theorem}
\noindent{\bf Proof:} Please refer to the\tr. \done

Based on Theorem~\ref{theo:error_bound}, we see that
Algorithm~\ref{alg:alg1}  can estimate $h^L(j,S)$ with a maximum error
$\epsilon$  with least probability $1-\delta$ ($0\!<\!\delta, \epsilon
\!<\!1$) by setting $R \geq \log(2L/\delta)/(2(1-c)^2{\epsilon}^2)$.

\section{Centrality Maximization}\label{sec:maximization}

In this section, we develop efficient algorithms to address the centrality
maximization problem \CP defined in \S\ref{subsec:centrality_def}. Noted
that even though we can efficiently estimate the decayed hitting probability
$h(j,S)$ by using random walks (see \S\ref{sec:computation}), finding a set
$S$ of $k$ nodes in a SAN to maximize its influence centrality $I(S)$ is
still computationally difficult as it requires to estimate the influence
centrality of all combinations of $k$ nodes. In particular, \CP is NP-hard.

\begin{theorem}
The centrality maximization problem \CP  is NP-hard.  \label{theo:NP}
\end{theorem}
\noindent{\bf Proof:} Please refer to the\tr. \done

To solve the centrality maximization problem \CP,  we develop
greedy-based approximation algorithms by exploiting the submodularity
property of $I(S)$. Specifically, we first show the submodularity property
and present a baseline greedy algorithm to maximize $I(S)$, and then develop
two novel optimization techniques to accelerate the greedy algorithm.

\subsection{Baseline Greedy Algorithm}

Before presenting the greedy-based approximation algorithm for maximizing
$I(S)$, we first show that $I(S)$ is a non-decreasing submodular function,
and the result is stated in the following theorem.

\begin{theorem}
The  centrality $I(S)$ is a non-decreasing submodular function.
\label{theo:submodularity}
\end{theorem}

\noindent{\bf Proof:} Please refer to the\tr.\done

Based on the submodularity property, we develop a greedy algorithm for
approximation when maximizing $I(S)$, and we call it {\em the baseline greedy algorithm}.
Algorithm~\ref{alg:baseline_greedy} describes this procedure.
To find a set of $k$ nodes to maximize $I(S)$, the algorithm works for $k$
iterations. In each iteration, it selects the node which maximizes the
increment of $I(S)$.

\begin{algorithm}[!t]
    \caption{Baseline Greedy Alg. for Maximizing $I(S)$}
    \footnotesize
    \begin{algorithmic}[1]
    \Require A hypergraph, and a parameter $k$;
    \Ensure A set $S$ of $k$ nodes for maximizing $I(S)$;
    \State  $S$ $\leftarrow \emptyset$, $I(S)\leftarrow 0$;
    \For{ $s = 1$ to $k$}
    	\For{$u \in (V-S)$}
            \State $I(S\cup\{u\}) \leftarrow 0$;
    		\For{$j \in (V-S \cup \{u\})$} \label{line:mg1}
    			\State $I(S\cup\{u\}) \leftarrow I(S\cup\{u\}) + h(j,S\cup \{u\})$;\label{line:mg2}
    		\EndFor
		\EndFor
        \State $v\leftarrow \arg \max_{u\in(V-S)}{I(S\cup\{u\})-I(S)}$;
		\State $S \leftarrow S \cup \{v\}$;\
	\EndFor
\end{algorithmic}
\label{alg:baseline_greedy}
\end{algorithm}

Recall that the time complexity for estimating the influence of a set $S$ to
a particular node $j\notin S$, i.e., $h(j,S)$, is $O(RL)$ (see
\S\ref{subsec:monte_carlo}). Thus, the total time complexity for the
baseline greedy algorithm is $O(kn^2RL)$ where $n$ denotes the total number
of users in the SAN, because estimating the influence of a set $S\cup \{u\}$
requires us to sum up its influence to all nodes,  and we need to check
every node $u$ so as to select the  one which maximizes the increment of
$I(S)$. Although the baseline  greedy algorithm gives a polynomial time
complexity, it is  inefficient when the number of users becomes large. To
further speed up the computation, we present two novel optimization
techniques in the next subsection.

\subsection{Optimizations} \label{subsec:optimization}

\noindent $\bullet$ {\bf Parallel Computation}: The key component in the
      greedy algorithm is to measure the  marginal increment of the
      influence  after adding node $u$, i.e.,
      $\Delta(u)=I(S\cup \{u\})-I(S)$, which can be  derived as
      follows.
       \begin{eqnarray*}
      \small
         \Delta(u) = \left[1-\sum\nolimits_{h=1}^{\infty}c^{h}P(u,S,h)\right]\times
          \left[1+\sum\nolimits_{j \in (V-S\cup\{u\})} \sum\nolimits_{h=1}^{\infty}c^{h}P^{S}(j,\{u\},h)\right].
      \end{eqnarray*}

\noindent In the baseline greedy algorithm, $\Delta(u)$'s are  computed
      sequentially, which as a result incurs a  large time overhead.
      Our main idea to speed up the computation is to estimate the
      marginal increment of all nodes, i.e., $\Delta(u)$ for every $u$, {\em in
      parallel}. Specifically, when performing $R$ random walks from a
      particular node $j$, we measure the contribution of $j$ to the
      marginal increment of every node. In other words, we obtain
      $P^{S}(j,u,h)$ for every $u$ by using only the $R$ random walks
      starting from $j$. As a result, we need only $O(nR)$ random walks to
      derive the marginal increment of all nodes, i.e., $\Delta(u)$ for
      every $u$, instead of $O(n^2R)$ random walks as in the baseline
      greedy algorithm.

\noindent
$\bullet$ {\bf Walk Reuse}: The core idea  is that in each iteration of
       choosing one node to maximize the marginal increment, we record the
       total $O(nR)$ random walks in memory, and apply the updates
       accordingly after one node is added into the result set. By doing
       this, we can reuse the $O(nR)$ random walks to derive the marginal
       increment in the next iteration instead of starting new random
       walks from each node again.

\begin{footnotesize}
\begin{algorithm}[!t]
    \caption{Optimized Greedy Algorithm}
    \footnotesize
    \label{alg:optimized}
    \begin{algorithmic}[1]
    \Require A hypergraph and a parameter $k$;
    \Ensure A set $S$ of $k$ nodes for maximizing $I(S)$;
    \State  $ S \leftarrow \emptyset$, $score[1...n]\leftarrow 0$, $P[1...n]\leftarrow 0$;
    \For{$j \in V$}
        \For{$r$ = 1 to $R$}
            \State $i \leftarrow j$, $visited \leftarrow \emptyset$;
            \For{$t$ = 1 to $L$}
                \State $visited \leftarrow visited \cup \{i\}$;
                \State $i \leftarrow$ Select a user according to the transition prob.;
                \State $RW[j][r][t] \leftarrow i$;
                \If{$i \notin visited$}
                    \State $index[i].add(item(j,r,t))$;
                    \State $score[i] \leftarrow score[i]+\frac{c^t}{R}$;
                \EndIf
            \EndFor
    	\EndFor
    \EndFor
    \State $v \leftarrow \arg \max_{u \in V}score[u]$;
	\For{ $s$ = 2 to $k$}
        \State Update ($RW, index, P, score, S, v, L$), $S \leftarrow S \cup \{v\}$;
        \State $v \leftarrow \arg \max_{u \in (V-S)}(1-P[u])(1+score[u])$;
   \EndFor
   \State $S \leftarrow S \cup \{v\}$;
   \vskip -20pt
\end{algorithmic}
\end{algorithm}

\begin{algorithm}[!t]
\footnotesize
    \caption{Update Function}
    \label{alg:update}
    \begin{algorithmic}[1]
    \Function{Update ($RW, index, P, score, S, v, L$)}{}
        \For{$w \in index[v]$}
            \State $k \leftarrow L$;
            \For{$t=1$ to L}
                \If{$RW[w.j][w.r][t]\in S$}
                    \State $k \leftarrow t$;
                    \State break;
                \EndIf
            \EndFor
            \If{$k==L$}
                \State $P[w.j] \leftarrow P[w.j]+c^{t}/R$
            \EndIf
            \For{$i=w.t+1$ to $k$}
               \State $u \leftarrow RW[w.j][w.r][i]$, $score[u] \leftarrow score[u]-c^{t}/R$;
            \EndFor
    	\EndFor
    \EndFunction
\end{algorithmic}
\end{algorithm}
\end{footnotesize}

By incorporating the above  optimization techniques, we can reduce the time
complexity to $O(nRL)$, where $L$ denotes the maximum walk length. In other words,
 we can use the $L$ leading terms to estimate
$\sum_{h=1}^{\infty}c^{h}P^{S}(j,\{u\},h)$ and
$\sum_{h=1}^{\infty}c^{h}P(u,S,h)$ as described in \S\ref{sec:computation}.
Thus, we let each  walk runs for $L$ steps at most.
Algorithm~\ref{alg:optimized} states the procedure. We use $score[u]$ and $P[u]$
 to record $\sum_{j \in V-S\cup\{u\}}
\sum_{h=1}^{\infty}c^{h}P^{S}(j,\{u\},h)$ and
$\sum_{h=1}^{\infty}c^{h}P(u,S,h)$ for computing $\Delta(u)$, respectively.
Algorithm~\ref{alg:optimized} runs in two phases. The first phase (line
1-13) is to select the first seed node by running random walks and also
record all the walking information for reuse.
 The second phase (line 14-18) is  to select the remaining $k-1$ nodes based on the stored  information
which requires to be updated after selecting each node. We give the update
function in Algorithm~\ref{alg:update}.

The update function is to update the walk information stored in $score$ and
$P$. Every time after we selecting a  node $v$,  the random walk in the
following iterations should stop when it encounters $v$,  and the values
stored in $score$ and $P$ should change accordingly. To achieve this, for
each random walk that hits $v$ (line 2), we first check if it has visited
any node in $S$  (line 4-7). If not,  we increase $P[w.j]$ after adding $v$
in $S$ (line 8,9). Since the following walks should stop when hitting $v$,
we  update $score[u]$  if node $u$ is visited after $v$ (line 10-12).

\section{Experiments}\label{sec:experiment}
To show the efficiency and effectiveness of our approach, we conduct
experiments on  real-world datasets. In particular, we first show that
incorporating online activities in seed selection can lead to a significant
improvement on the influence spread, i.e., influence more users with the
same seed size. Then we show that our IM-RW algorithm takes much less
running time  than the state-of-the-art influence maximization algorithm,
while achieves almost the same influence spread.

\vspace{-5pt}
\subsection{Datasets}

 We consider three datasets from social rating
systems: Ciao \cite{Tang-etal12b}, Yelp \cite{yelpdataset} and Flixster
\cite{Jamali10_flixster}. Such social rating networks are composed of a
social network, where the links can be interpreted as either friendships
(undirected link) or a following relationship (directed link), and a rating
network, where a link represents that a user assigns a rating (or writes a
review) to a product. Assigning  a rating corresponds to an online activity,
and multiple users assigning ratings to the same product means that they
participate in the same online activity. In the rating network, we remove
rating edges if the associated rating is less than 3 so as to filter out the
users who dislike a product. Through this we guarantee that all the
remaining users who give ratings to the same product have similar interests,
e.g., they all like the product. Since the original Flixster dataset is too
large to run the state-of-the-art influence maximization algorithms, we
extract only a subset of the Flixster dataset for comparison studies. In
particular, since the OSN of Flixster is almost a connected component, we
randomly select a user, and run the breadth-first search algorithm until we
get 300,000 users. We state the  statistics of the three datasets in
Table~\ref{tab:dataset_info}.  All algorithms are run on a server with two
Intel Xeon E5-2650 2.60GHz CPU and 64GB memory.

\begin{table}[!h]
  \centering
  \begin{adjustbox}{max width=8.5cm}
  \begin{tabular}{*{7}{|c}|}
  \hline
 Dataset & Users & Links in OSN & Products & Ratings &  OSN Type \\
   \hline
Ciao & 2,342 & 57,544 & 15,783 & 32,783 & directed \\
 \hline
Yelp & 174,100 & 2,576,179 & 56,951 & 958,415 & directed \\
 \hline
Flixtser & 300,000 & 6,394,798 & 28,262 & 2,195,134 & undirected \\
 \hline
\end{tabular}
\end{adjustbox}
  \caption{ Datasets Statistics.}
  \vskip -20pt
  \label{tab:dataset_info}
\end{table}

\subsection{The Benefit of Incorporating  Activities}

\label{sec:MeritAct}

We first show that incorporating  online activities in seed selection can
lead to a significant improvement on the influence spread. We fix the seed
size $k$  as 50. To show the impact of activities, we use the
state-of-the-art influence maximization algorithm IMM
\cite{tang2015influence} to select the seed set on OSNs and use our IM-RW
algorithm to select the seed set on SANs which take online activities into
account.
Then we use simulations to estimate the expected influence spread of the
selected $k$ users on SANs
and denote the results as $\sigma(\mbox{OSN})$ and $\sigma(\mbox{SAN})$,
respectively. Finally, we define the improvement ratio on the expected
influence spread as $[\sigma(\mbox{SAN})-\sigma(\mbox{OSN})] /
\sigma(\mbox{OSN})$.

To present the key insights, we consider the simple case in which there is
only one type of users and online activities. Namely, all users have a same
value of $\alpha$ which  indicates the weight of activities. We emphasize
that our model also works in the general case of multiple types of users and
online activities.

\begin{figure*}[!ht]
\centering
\begin{tabular}{c@{\quad}c@{\quad}c}
\includegraphics[width=0.3\linewidth]{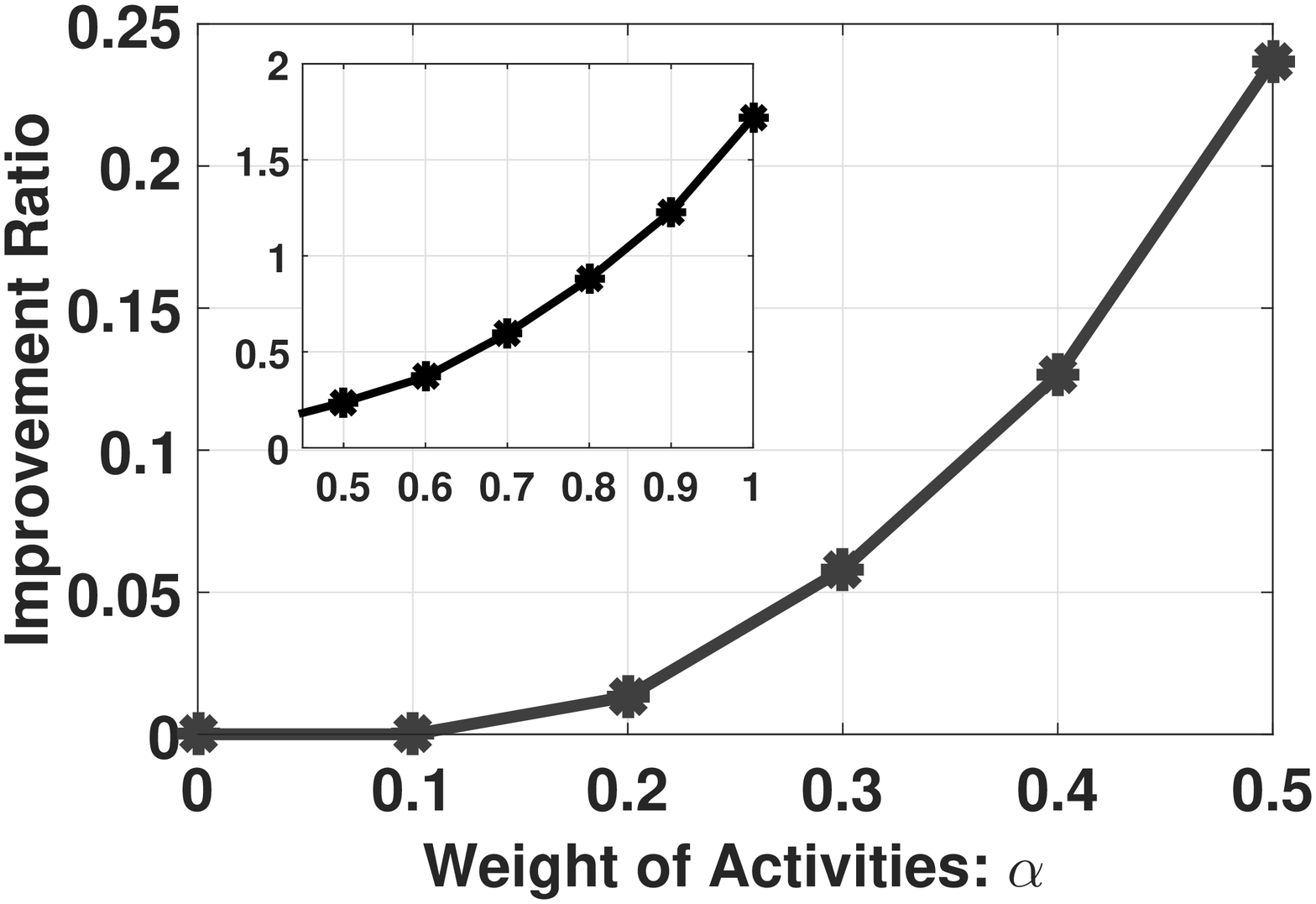}
\label{Fig:ImpactActCiao}
&
\includegraphics[width=0.3\linewidth]{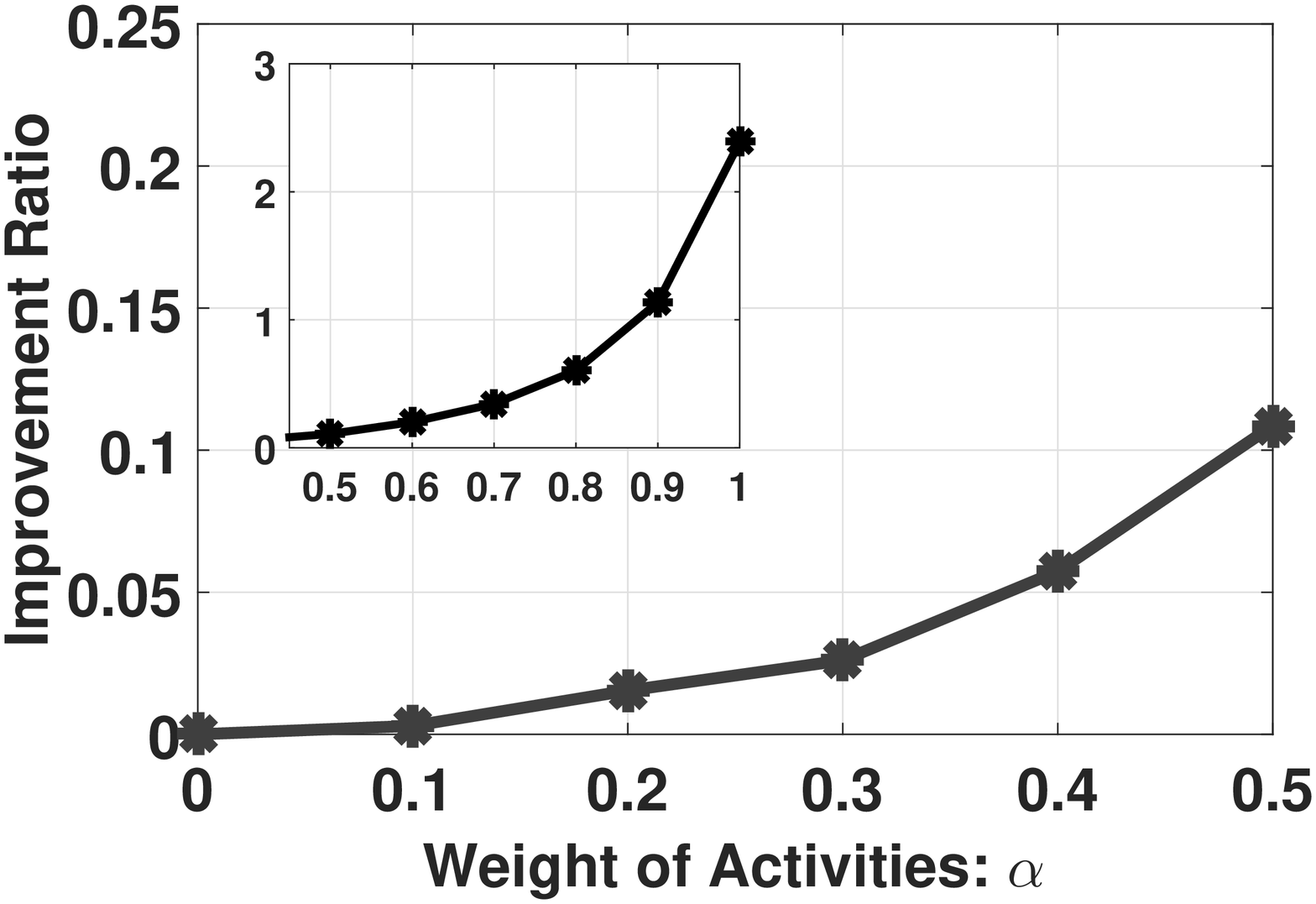}
\label{Fig:ImpactActYelp}
&
\includegraphics[width=0.3\linewidth]{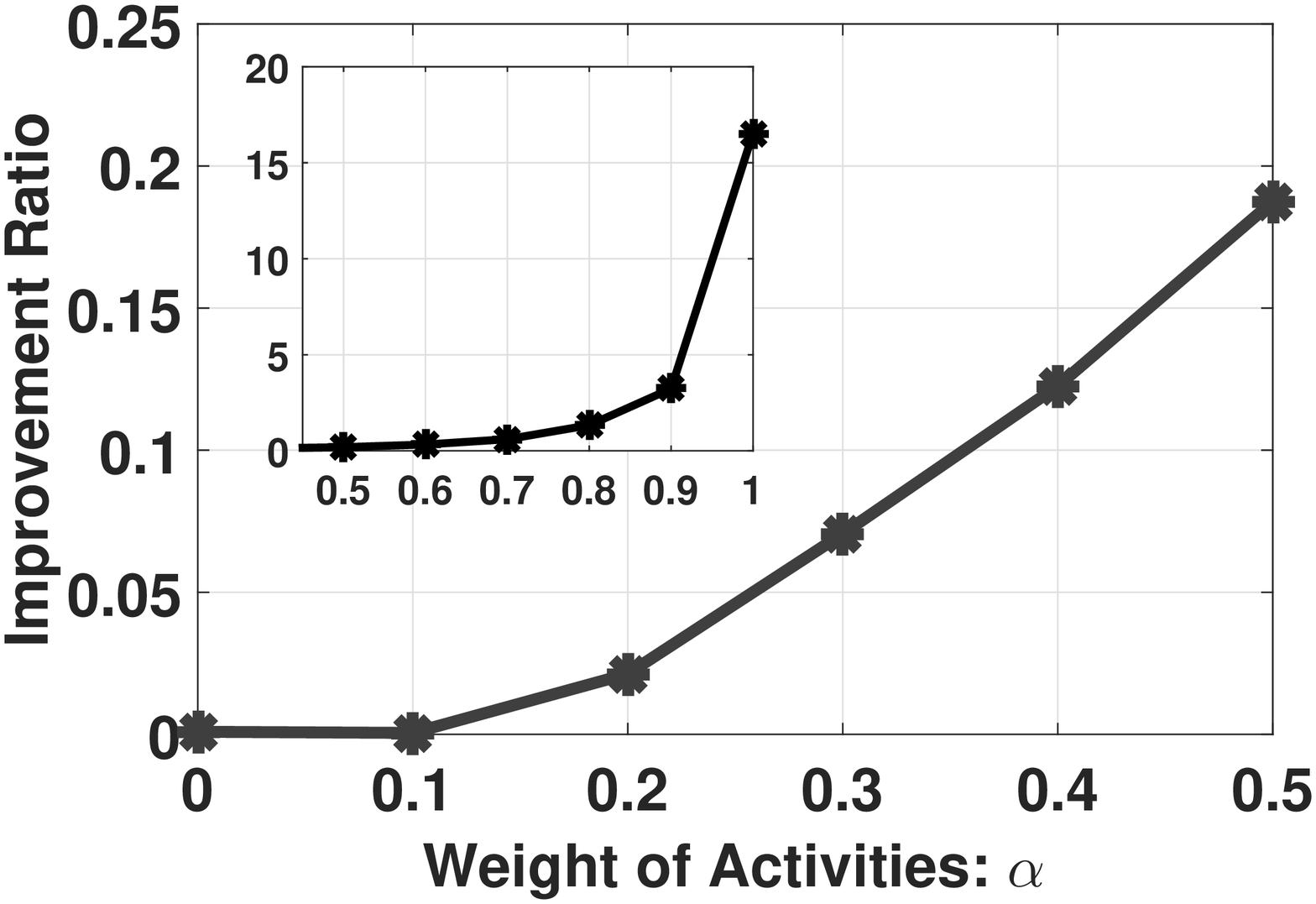}
\label{Fig:ImpactActFlixster}
\\
\mbox{\small (a) Ciao } &
\mbox{\small (b) Yelp }    &
\mbox{\small (c) Flixster }
\end{tabular}
\vspace{-5pt}
\caption{Impact of online activities on influence spread.}
\vspace{-20pt}
\label{fig:impact_spread}
\end{figure*}

Figure \ref{fig:impact_spread} depicts the improvement of incorporating
online activities by varying the weight of activities $\alpha$ from 0 to 1.
The horizontal axis shows the value of $\alpha$, and the vertical axis
presents the corresponding improvement ratio.  From
Figure~\ref{fig:impact_spread}, one can observe that the improvement ratio
is 0 when $\alpha=0$. This is because users are not affected by other users
through online activities when $\alpha=0$.
As $\alpha$ increases, the improvement ratio also increases. This shows that
as users are more prone to be affected by other users through online
activities, incorporating online activities bring larger benefit. When
$\alpha=0.5$, the improvement ratio is around 25\% for Ciao dataset. That
is, we can influence 25\% more users when incorporating  online activities
in the seed selection.  Similar conclusions can also be observed for the
datasets of Yelp and Flixster.
 It is
interesting to observe that as $\alpha$ approaches to one, the improvement
ratio reaches up to 16 for Flixster, which implies  a more than an order of
magnitude  improvement.  In summary, incorporating online activities in the
seed selection by using IM-RW significantly improves the selection accuracy.

\subsection{Performance Evaluation of IM-RW}

\begin{figure*}[!tb]
\centering
\begin{tabular}{c@{\quad}c@{\quad}c}
\includegraphics[width=0.3\linewidth]{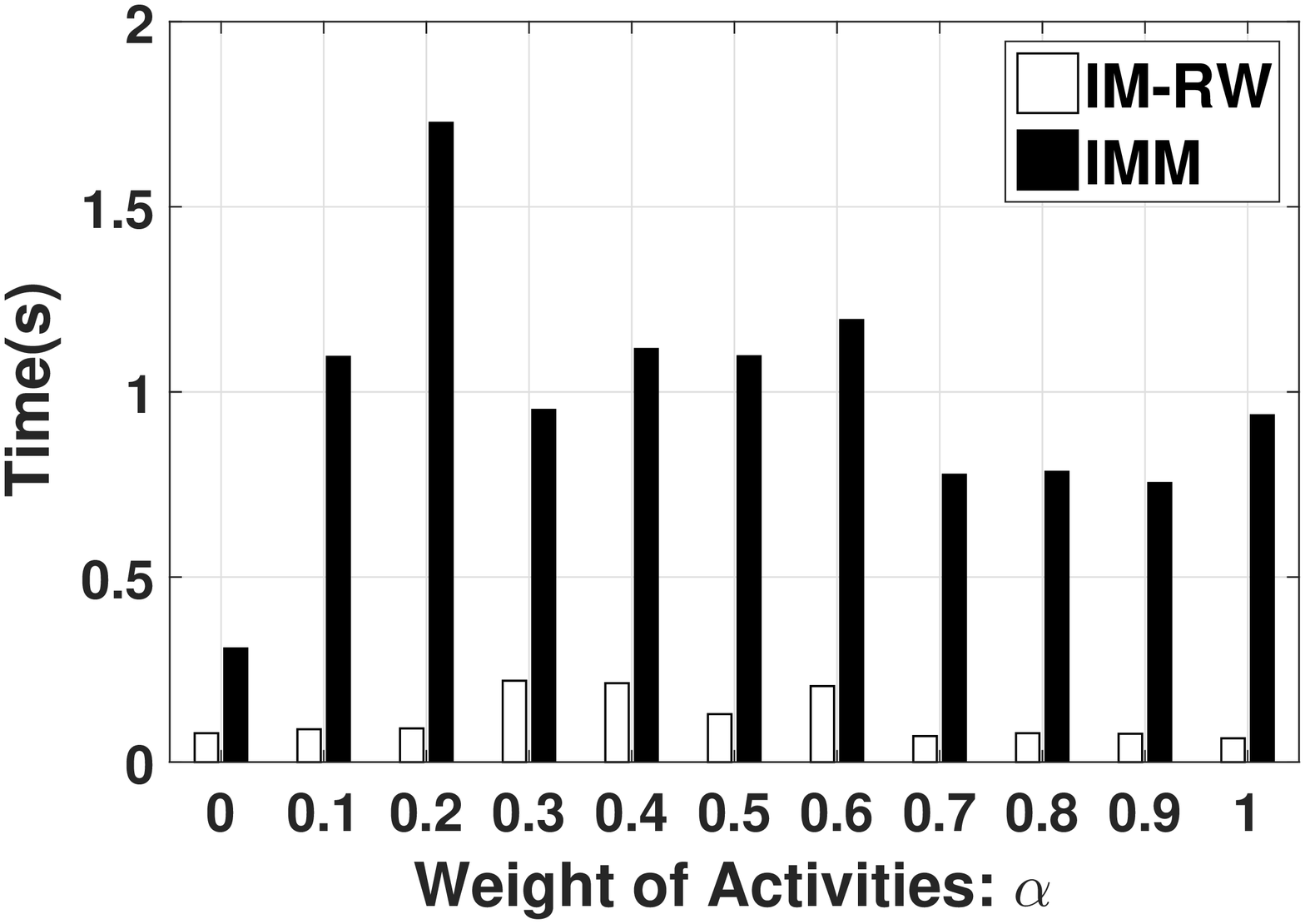} &
\includegraphics[width=0.3\linewidth]{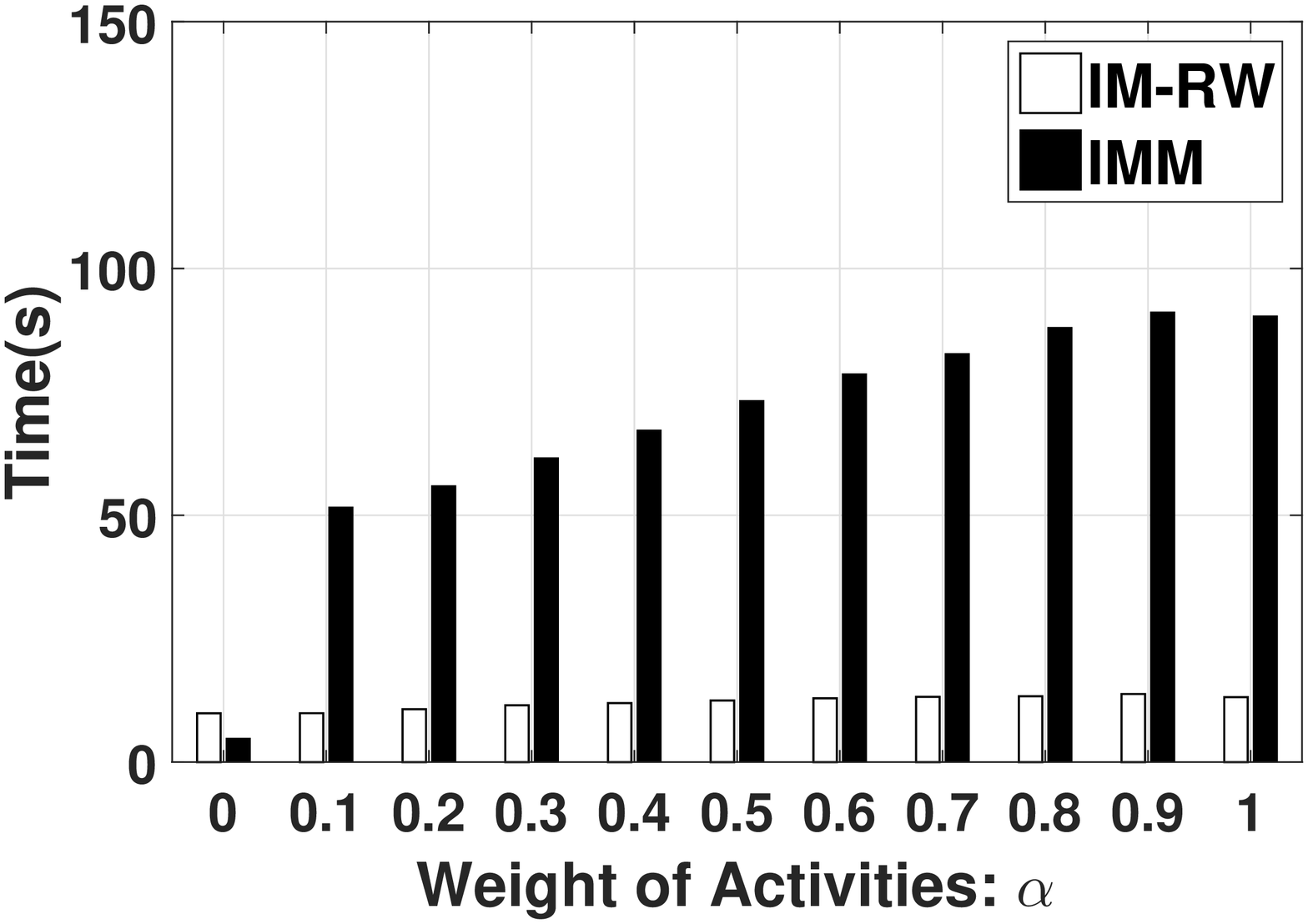} &
\includegraphics[width=0.3\linewidth]{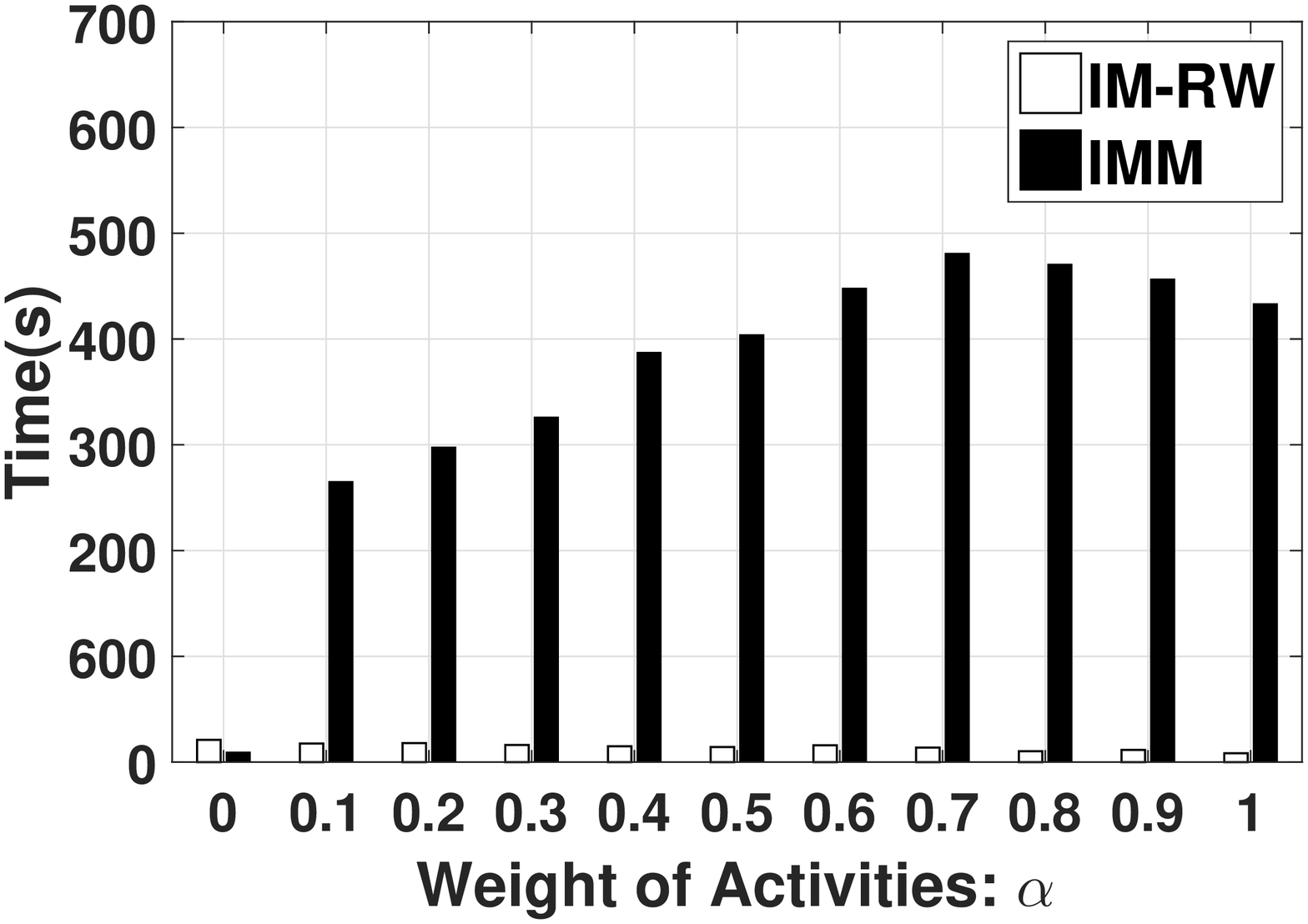} \\
\mbox{\small (a) Ciao } &
\mbox{\small (b) Yelp }    &
\mbox{\small (c) Flixster }
\end{tabular}
\vspace{-5pt}
\caption{Running time of IM-RW and IMM with different activity weights.}
\label{fig:time_algorithm}
\vspace{-10pt}
\end{figure*}

\begin{figure*}[!tb]
\centering
\begin{tabular}{c@{\quad}c@{\quad}c}
\includegraphics[width=0.3\linewidth]{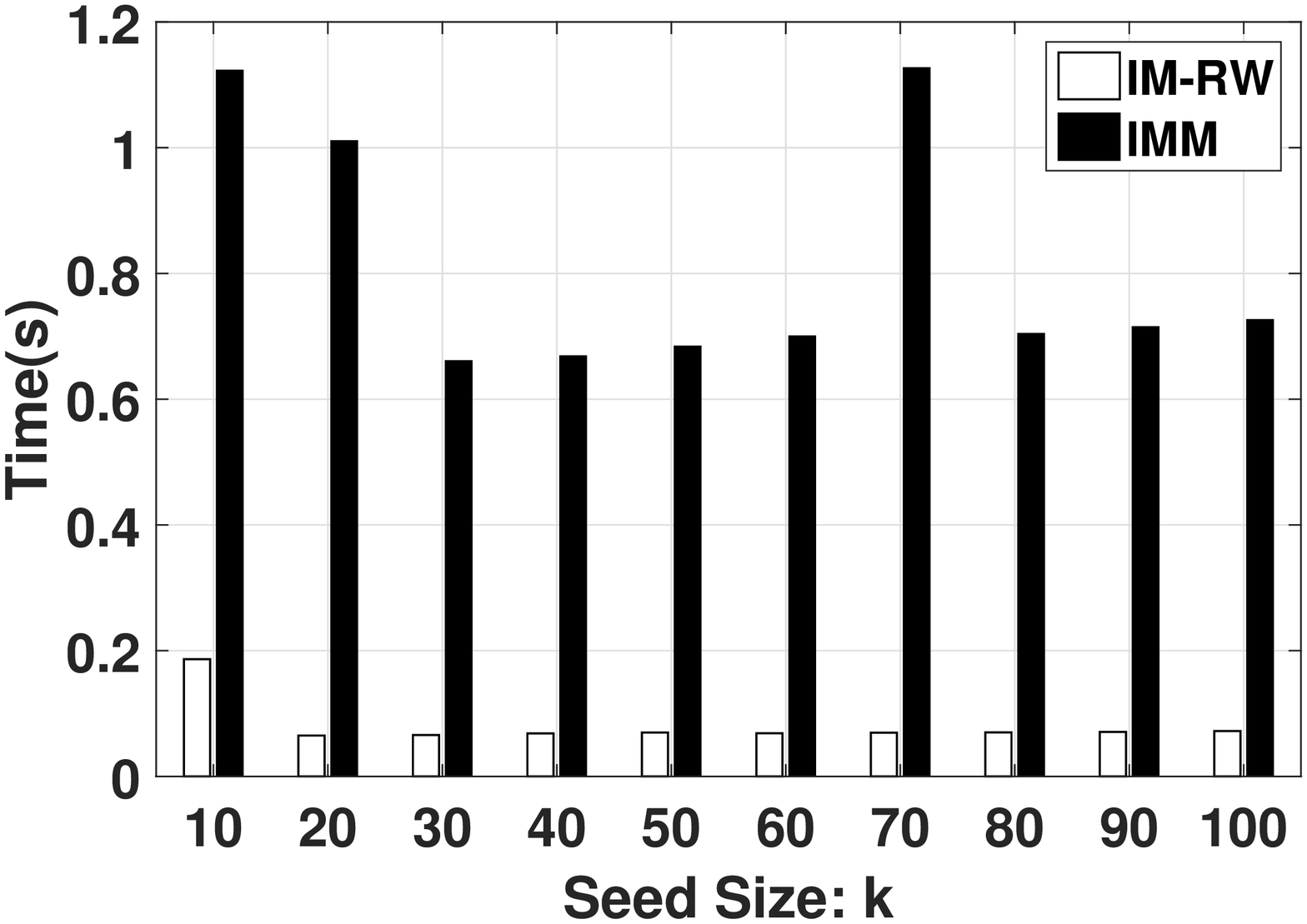} &
\includegraphics[width=0.3\linewidth]{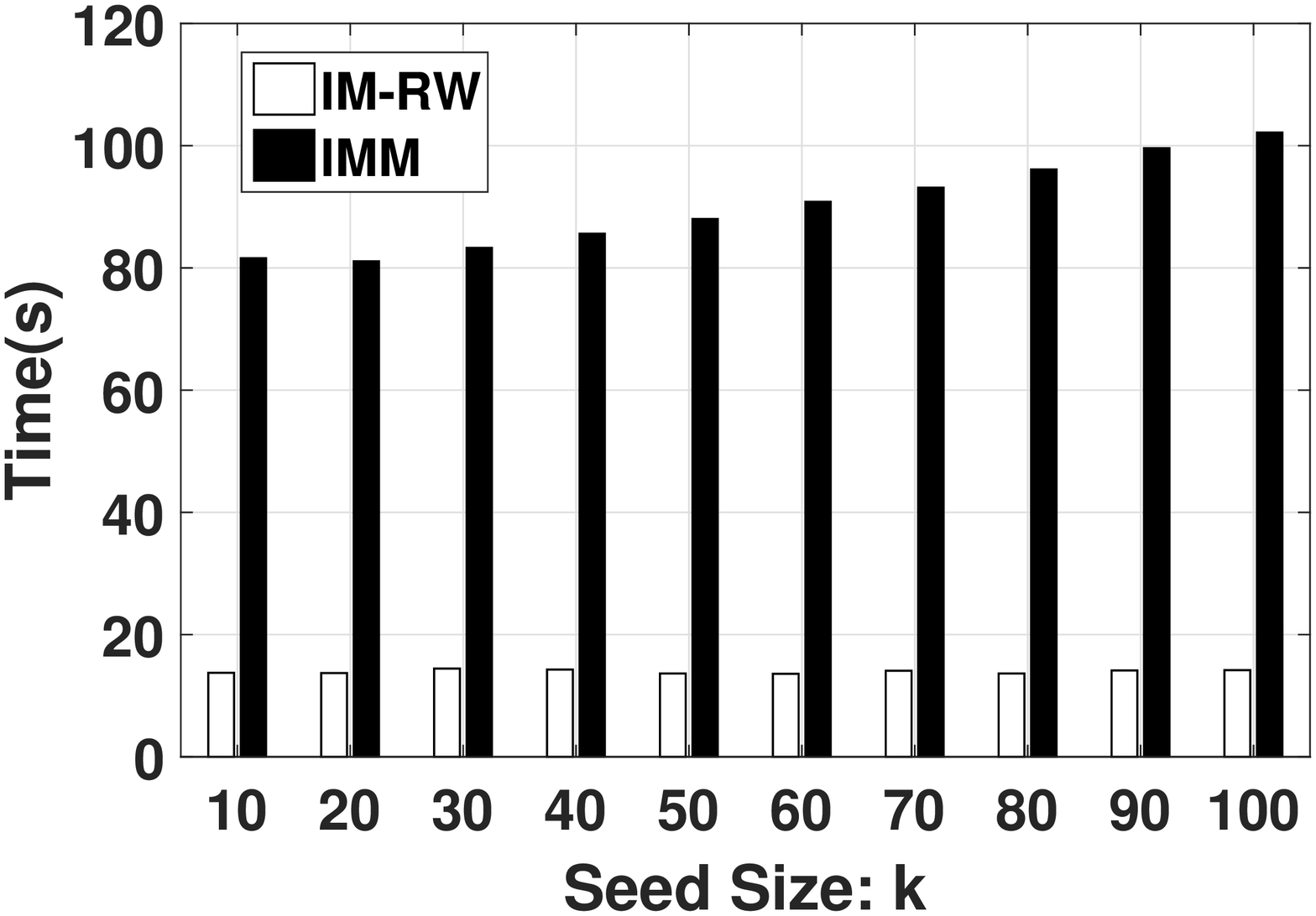} &
\includegraphics[width=0.3\linewidth]{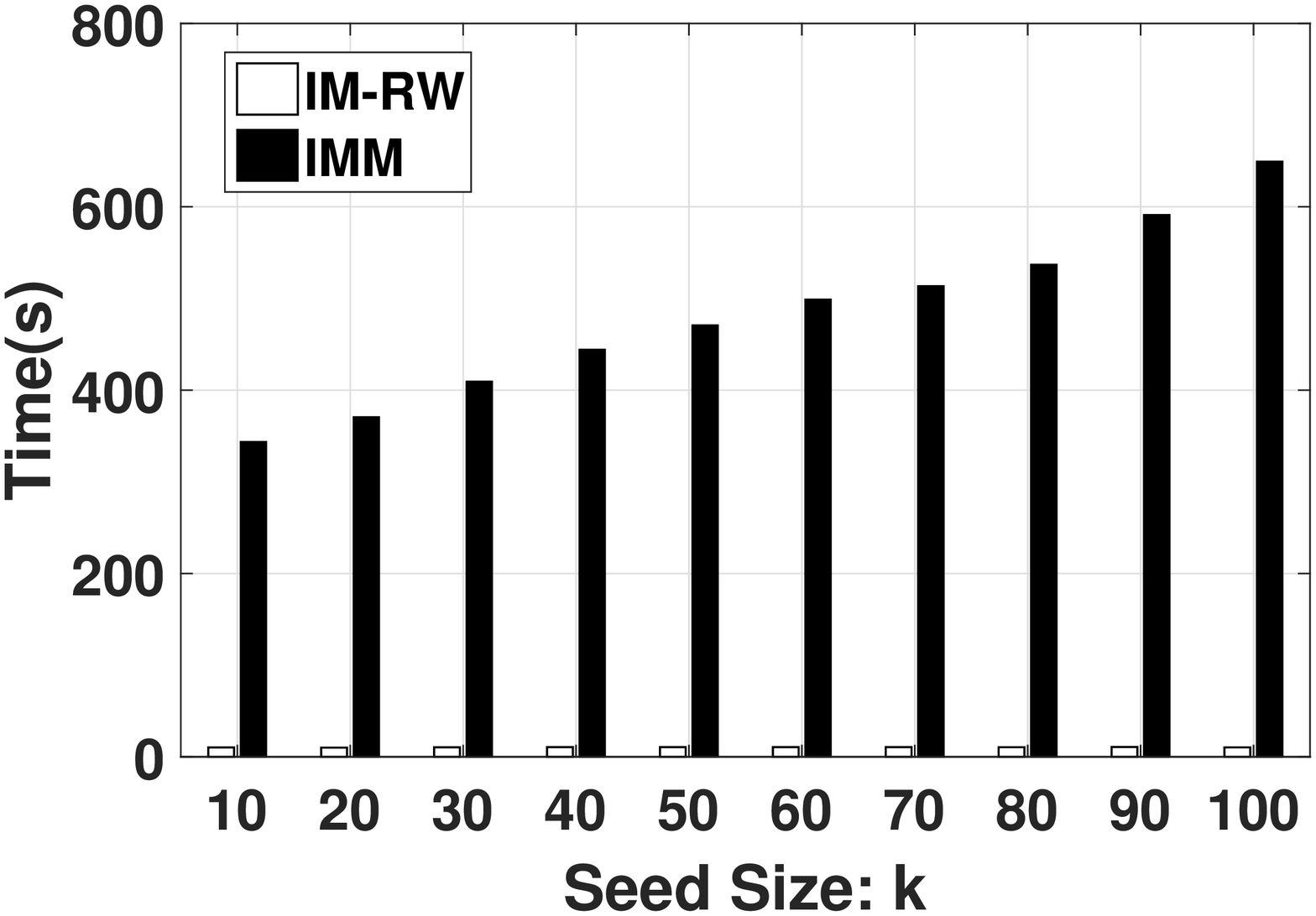} \\
\mbox{\small (a) Ciao } &
\mbox{\small (b) Yelp }    &
\mbox{\small (c) Flixster }
\end{tabular}
\vspace{-5pt}
\caption{Running time of IM-RW and IMM with different seed sizes.}
\vspace{-20pt}
\label{fig:time_k}
\end{figure*}

In this subsection, we validate the efficiency and effectiveness of IM-RW by
comparing it with  IMM, which is the state-of-the-art algorithm for solving
influence maximization problem in OSNs, from two aspects, the running time
and the influence spread. Note that IMM was originally developed for OSNs
without online activities being considered, so for fair comparison, we
transform SANs to a weighted graph by also taking online activities into
account, and then apply IMM on the weighted graph to derive the most
influential nodes.

We first compare the running time of IM-RW and IMM by varying  the weight of
activities $\alpha$ and the seed size $k$, and the results are presented in
Figure~\ref{fig:time_algorithm} and Figure~\ref{fig:time_k}. Specifically,
Figure~\ref{fig:time_algorithm} shows that IMM takes much longer time than
IM-RW, especially when the network is large and online activities become
more important (i.e., with larger $\alpha$). This is because as $\alpha$
increases, the time cost of IMM depends more on user-activity-user links
than user-user links. Thus, as the amount of user-activity-user links is
much more than that of user-user links in SANs,  the time cost of IMM will
increase. On the other hand, when we fix $\alpha$ as 0.8 and vary the seed
size $k$, Figure~\ref{fig:time_k} also shows  that IMM takes much longer
time than IM-RW under all settings. Therefore, we can conclude that our
IM-RW algorithm really improves the efficiency of solving the influence
maximization problem in SANs with online activities being considered.

\begin{figure*}[!tb]
\centering
\begin{tabular}{c@{\quad}c@{\quad}c}
\includegraphics[width=0.3\linewidth]{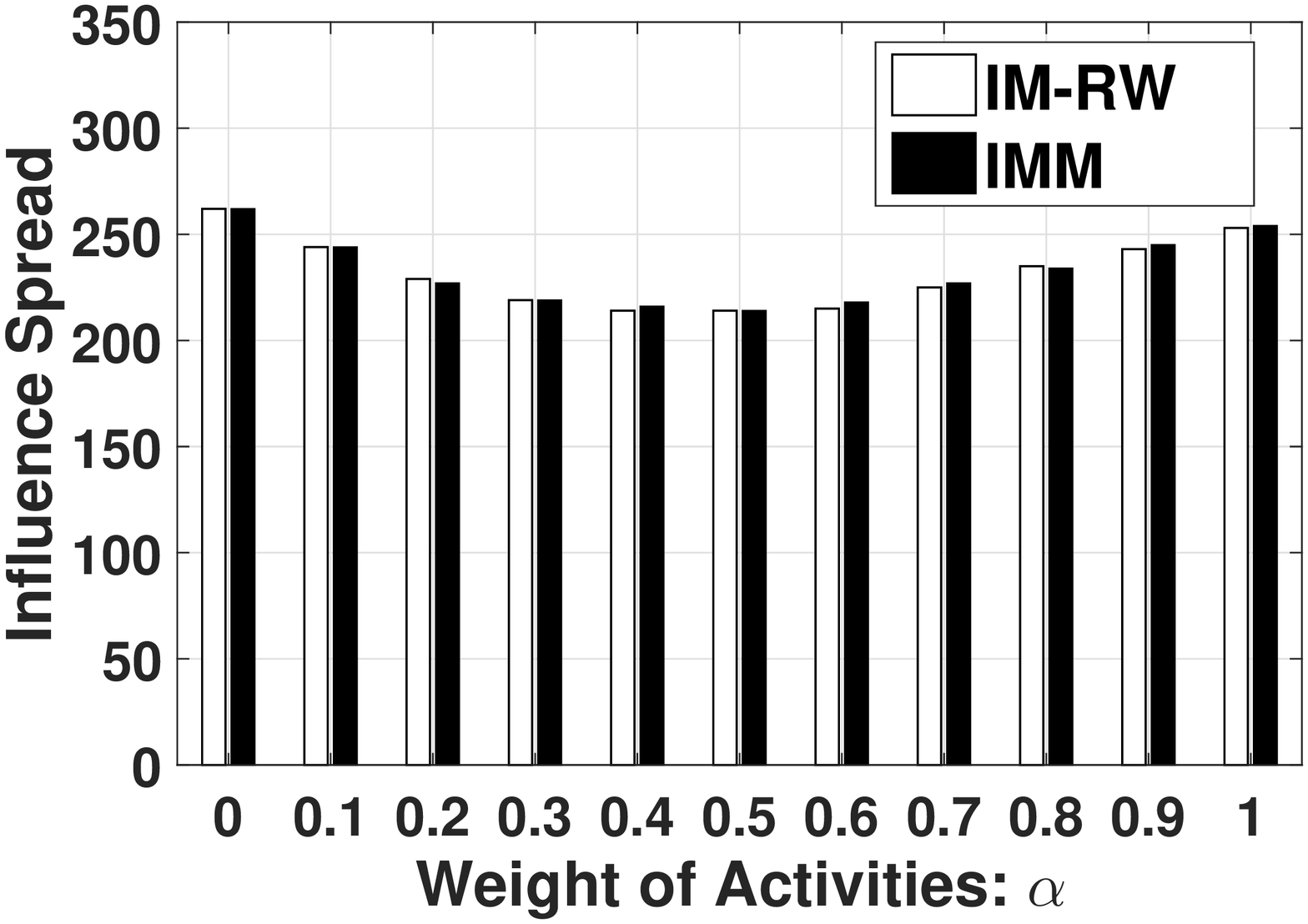} &
\includegraphics[width=0.3\linewidth]{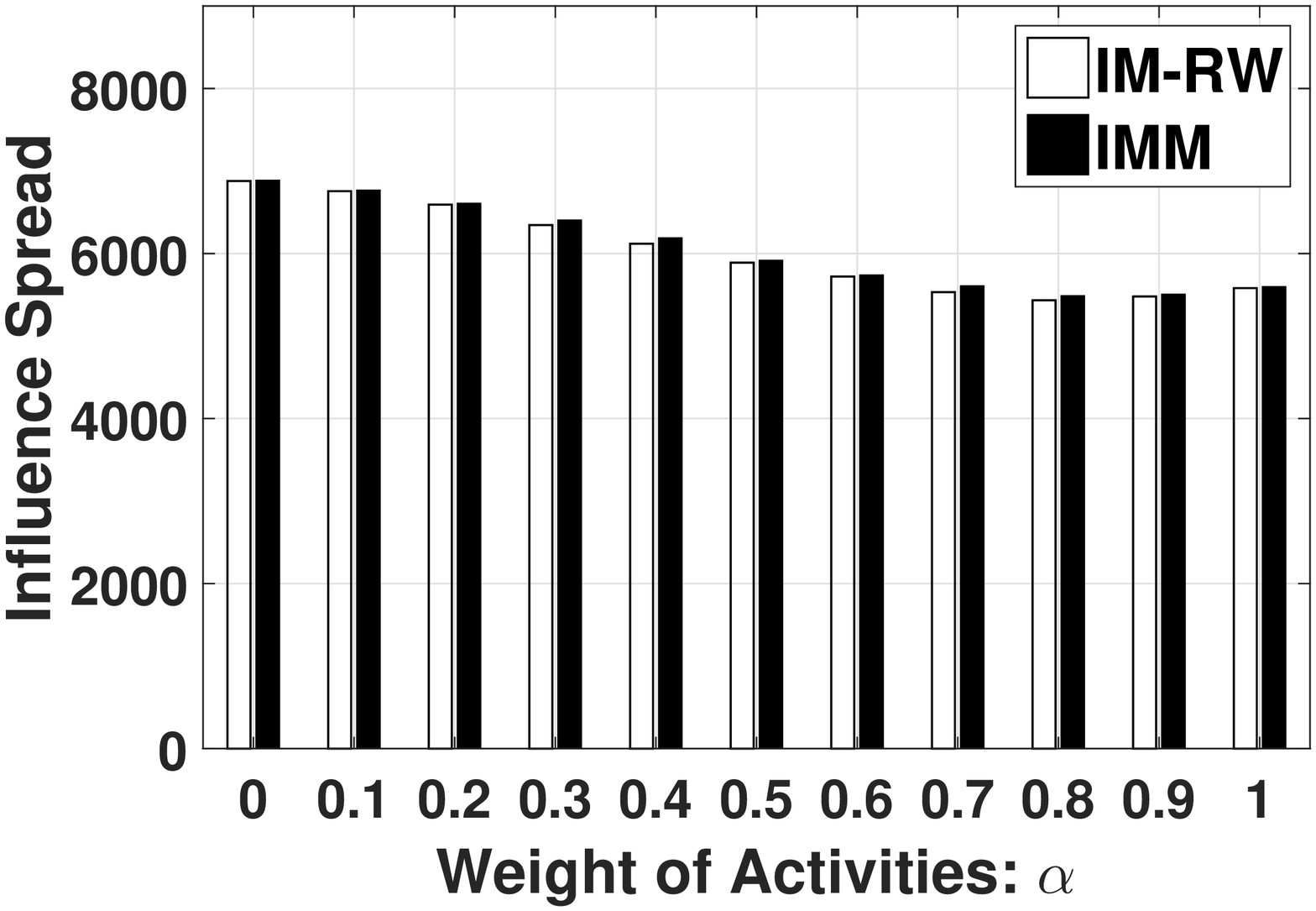} &
\includegraphics[width=0.3\linewidth]{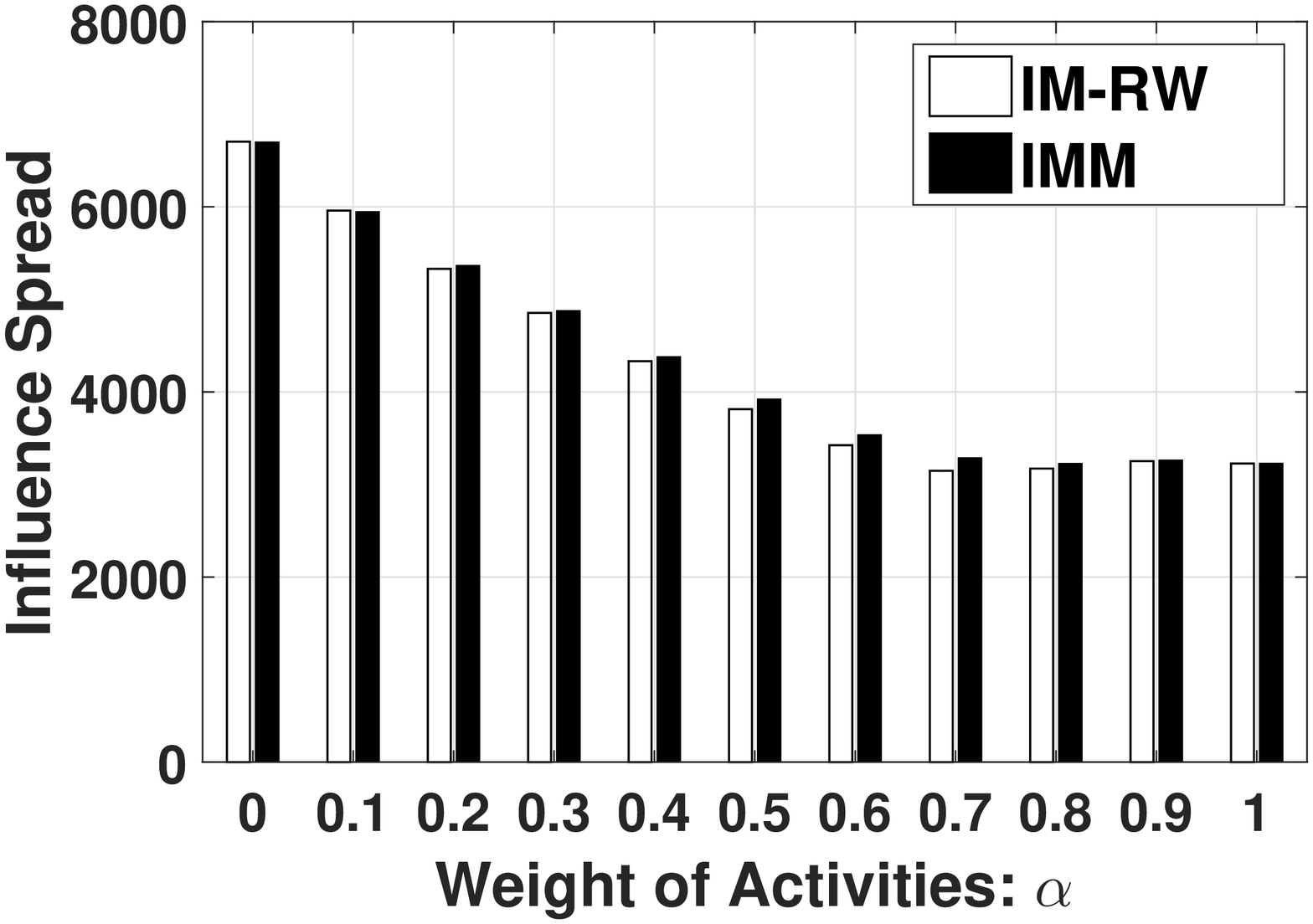} \\
\mbox{\small (a) Ciao } &
\mbox{\small (b) Yelp }    &
\mbox{\small (c) Flixster }
\end{tabular}
\vskip -5pt
\caption{Influence spread of IM-RW and IMM.}
\label{fig:spread_algorithm}
\vspace{-20pt}
\end{figure*}

We further show the influence spread of the most influential users selected
by the two algorithms in Figure~\ref{fig:spread_algorithm}. The horizontal
axis shows the values of $\alpha$, and the vertical axis represents the
corresponding influence spread. We see that by taking online activities into
consideration, both IMM and IM-RW can achieve almost the same performance.
Because IMM is an influence maximization algorithm with theoretical
performance guarantees, we can conclude  that our IM-RW approach also has a
good performance to maximize the influence spread.

\noindent {\bf Summary:} Our IM-RW algorithm achieves a good performance in
both the running time and the influence spread by taking online activities
into account in SANs. In particular, comparing to the state-of-the-art
algorithm IMM,
our IM-RW algorithm achieves almost the same performance in seed selection,
while it only requires much less running time.

\section{Related Work} \label{sec:related}

Influence maximization problem in OSNs was first formulated by Kempe et al.
\cite{kempe2003maximizing},  and in this seminal work,  the authors proposed
 the IC model and the LT model.
Since then, this problem receives a lot of interests in academia in the past
decade \cite{chen2009efficient, chen2010scalable, chen2010scalableICDM}.
Because of the   NP-hardness under both the IC model \cite{chen2010scalable}
and the LT model \cite{chen2010scalableICDM}, many of the previous studies
focus on how to reduce the time complexity. Recently, Borgs et al.
\cite{Borgs2014Maximizing}   developed an algorithm which maintains the
performance guarantee while reduces the time complexity significantly, and
Tang et al. \cite{tang2014influence, tang2015influence} further improved the
method and proposed the IMM algorithm, which is the state-of-the-art
solution for influence maximization in OSNs.
Besides reducing the computation overhead,
several works improved the influence models, for example,
topic-aware influence model\cite{barbieri2012topic},
competitive influence model\cite{lin2015analyzing},
opinion-based influence model\cite{galhotra2016holistic} etc.

Centrality measure based approach was also studied, for example,  the
studies \cite{freeman1979centrality, everett1999centrality,
zhao2014measuring, chen2009efficient} find the most influential nodes based
on degree centrality and closeness centrality. In terms of random walk, it
is widely used to analyze big graphs, e.g., PageRank computation
\cite{page1999pagerank}, graph sampling \cite{zhao2015tale}, and SimRank
\cite{kusumoto2014scalable} etc.

We would like to emphasize that our work differs from existing studies which
address the  traditional influence maximization problem, while we take
online activities into consideration. When we consider these online
activities, only considering user-user links alone may not trigger the
largest influence spread. Although we can also transform the
user-activity-user links to user-user links, the underlying graph may become
extremely dense so that traditional methods may not be efficient.

\section{Conclusions}\label{sec:conclusion}

In this paper, we address  the influence maximization problem in SANs with a
random walk approach. Specifically, we propose a general framework  to
measure the influence of nodes in SANs via random walks on hypergraphs, and
develop a greedy-based algorithm with two novel optimization techniques to
find the top $k$ most influential nodes in SANs by using random walks.
Experiments with real-world datasets show that our approach greatly improves
the computation efficiency, while keeps almost the same performance in seed
selection accuracy  compared to IMM, the state-of-the-art algorithm.

\section*{Acknowledgements}
This work was supported by National Nature Science Foundation of China
(61303048 and 61379038), and Anhui Provincial Natural Science Foundation
(1508085SQF214).

\newpage
\section*{Appendix}

{\noindent \bf Proof of Theorem 1: }  Based on the definition of $h(j,S)$ in
Equation~(\ref{eq:decayed_hitting_prob}), we can get
\begin{equation*}
h(j,S)=\sum_{h=1}^{\infty}c^{h}P(j,S,h), \mbox{  for } j \notin
S,
\end{equation*}
where $P(j,S,h)$ denotes  the probability  that a random walk starting from
$j$ hits a node in $S$ at the $h$-th step. Now if we denote $p_{iS}$ as the
probability that a random walk starting from $i$ hits a node in $S$ in one
step,  we have

{\small
\begin{eqnarray*}
  h(j,S) &=& \sum_{h=1}^{\infty}c^{h}P(j,S,h) =\sum_{h=1}^\infty{c^h}\sum_{i \notin S}(\bQ^{h-1})_{ji}p_{iS}, \\
   &=&\sum_{i \notin S}\sum_{h=1}^\infty{c^h}(\bQ^{h-1})_{ij}p_{iS}=\sum_{i \notin S}c(\bI-c\bQ)^{-1}_{ji}p_{iS}, \\
   &=&c\be_{j}^T(\bI-c\bQ)^{-1}\bQ'\be. \hspace{3.8cm}
\end{eqnarray*}
}

Note that the largest eigenvalue of $c\bQ$ is less than one, so by further expanding
the expression above with an infinite series, we can rewrite
$h(j,S)$ as follows.

\begin{eqnarray*}
h(j,S)=c\be_{j}^T\bQ'\be+c^2\be_{j}^T\bQ\bQ'\be+c^3\be_{j}^T\bQ^2\bQ'\be+\cdots.
\label{eq:infinit_series}
\end{eqnarray*}\done

\noindent{\bf Proof of Theorem 2:} Let $X_1$,...,$X_R$ be $R$ independent
random variables with $X_r \in [0,1]$ for all $r = 1,\cdots ,R$, and set
$T=(X_1+...+X_R )/R$.  According to Hoeffding's inequality, we have
$P\{|T-E(T)|>\epsilon\}\leq 2\exp(-2 {\epsilon}^2R)$. By applying this
inequality, we have

{\small
\begin{eqnarray*}
 && P\{|\hat h^L(j,S)-h^L(j,S)|>\epsilon\} \\
 &=& P\Big\{|\sum_{t=1}^L\frac{c^t}{R}\sum_{r=1}^{R}X_r{(t)}-\sum_{t=1}^Lc^tE[X{(t)}]|>\epsilon\Big\},\\
 &\leq&  P\Big\{\sum_{t=1}^L|\frac{c^t}{R}\sum_{r=1}^{R}X_r{(t)}-c^tE[X{(t)}]|>\epsilon\Big\}, \\
 &\leq&  \sum_{t=1}^LP\Big\{|\frac{c^t}{R}\sum_{r=1}^{R}X_r{(t)}-c^tE[X{(t)}]|>(1-c)c^t\epsilon\Big\}, \\
 &\leq&  2L\exp(-2(1-c)^2\epsilon^2R).\hspace{4cm} \done
\end{eqnarray*}
}

\vskip -20pt

 \noindent{\bf Proof of Theorem 3:} We first briefly introduce
the vertex cover problem. An instance of vertex cover problem is specified
by a graph $G(V,E)$ and an integer $k$, and asks there exists a vertex set
$S\subseteq V$ such that $|S|\leq k$ and for every $(i,j)\in E$,
$\{i,j\}\cap S \neq \emptyset$.

We now map our centrality maximization problem into
an instance of the vertex cover problem
by taking the same graph $G(V,E)$ and
asking whether there exists a vertex set $S$ such that $|S|\leq k$
and $I(S)\geq (n-k)\times c+k$.
We aim to show that $S$ is a vertex cover
if and only if $I(S)\geq (n-k)\times c+k$.
Assuming $S$ is a vertex cover.
Note that $h(i,S)=1$ if $i\in S$ and $h(i,S)=c$ otherwise.
Observe that for every $(i,j)\in E$, $\{i,j\}\cap S \neq \emptyset$.
This implies that $I(S)=(n-k)\times c+k$.
Suppose $S$ is not a vertex cover,
then there should be an edge $(u,v)$ which satisfies that
$\{u,v\}\cap S = \emptyset$.
A random walk from $u$ passing through $v$ and at last
arriving at $S$ will have length at least 2.
So $h(u,S) = \sum_{j\in N(u)/v}cp_{uj}h(j,S) + cp_{uv}h(v,S)$.
Due to $v\notin S$, $h(v,S)<1$.
Thus $h(u,S)<c$, which contradicts.
Therefore $S$ is a vertex cover.
\done

\noindent{\bf Proof of Theorem 4:} We first show the non-decreasing
property. Note that since $h(j,S)=1$ if $j\in S$, so we rewrite $I(S)$ as
follows.
\begin{equation*}
  I(S)=|S|+\sum\nolimits_{j \in (V-S)}h(j,S).
\end{equation*}
Now suppose that a user $u\notin S$ is added into the set $S$, then the
marginal increment of the influence centrality
$\Delta(u)=I(S\cup\{u\})-I(S)$ can be derived as {\small
\begin{eqnarray*}
\Delta(u) &=& \sum_{j\in V}h(j,S\cup\{u\}) -  \sum_{j \in V}h(j,S),\\
  &=& 1+\sum_{j \in (V-S\cup\{u\})}h(j,S\cup\{u\})-\sum_{j \in (V-S)}h(j,S),\\
  &=& 1-h(u,S)+\sum_{j \in (V-S\cup\{u\})} \Big[h(j,S\cup\{u\})-h(j,S)\Big].
\end{eqnarray*}
}
According to the definition of $h(j,S)$ in
Equation~(\ref{eq:infinit_series}) and the random walk interpretation, we
rewrite $h(j,S)$ as
\begin{equation*}
h(j,S)=\sum_{h=1}^{\infty}c^{h}P(j,S,h), \mbox{  for } j \notin
S,
\end{equation*}
where $P(j,S,h)$ denotes the probability that a random walk starting from
$j$ hits a node in $S$ at the $h$-th step for the first time. Now we can
rewrite  $h(j,S\cup\{u\})-h(j,S)$ as
{\small
\begin{eqnarray*}
& &h(j,S\cup\{u\})-h(j,S)\\
&=& \sum_{h=1}^{\infty}c^{h}P(j,S\cup\{u\},h) - \sum_{h=1}^{\infty}c^{h}P(j,S,h)\\
&=& \sum_{h=1}^{\infty}c^{h}\left[P^{\{u\}}(j,S,h)+P^{S}(j,\{u\},h)\right]- \sum_{h=1}^{\infty}c^{h}\left[P^{\{u\}}(j,S,h)+ P^{S}(j,\{u\},h)P(u,S,h)\right ] \\
&=& \sum_{h=1}^{\infty}c^{h}P^{S}(j,\{u\},h)\Big[1-\sum_{h=1}^{\infty}c^{h}P(u,S,h)\Big]\\
&=& \sum_{h=1}^{\infty}c^{h}P^{S}(j,\{u\},h)\Big[1-p(u,S,h)\Big],
\end{eqnarray*}
} where $P^T(j,S,h)$ represents the probability that a random walk starting
from $j$ hits a node in $S$ at the $h$-th step for the first time without
passing any node in $T$. Therefore,  $\Delta(u)$ can be derived as follows.
{\small
\begin{eqnarray}
\Delta(u)&=& I(S\cup\{u\})-I(S)\nonumber\\
&=& (1\!-\!h(u,S))\Big[1\!+\!\!\!\!\!\!
\sum_{j \in V-S\cup\{u\} }\!\sum_{h=1}^{\infty}\!c^{h}\!P^{S}(j,\{u\},h)\Big].
\label{eq:marginal_gain}
\end{eqnarray}
}
Note that $0<c<1$ and
$\sum_{h=1}^{\infty}P(u,S,h) \leq 1$, so we  have $h(u,S)\leq1$ and
$1-h(u,S) \geq 0$. That is,  $\Delta(u) \geq 0$, and $I(S)$ is a
non-decreasing function.
We now show that $I(S)$ is a submodular function. Mathematically, we only
need to prove that the  inequality $I(S\cup\{u\})-I(S) \geq
I(T\cup\{u\})-I(T), \mbox{ for } S \subseteq T$, holds.  Note that
$P^{S}(j,\{u\},h)\geq P^{T}(j,\{u\},h)$ if $S \subseteq T$. Besides,
according to the non-decreasing feature of $I(S)$, we have $h(u,S)\leq
h(u,T)$. 
Based on
 these
inequalities and Equation~(\ref{eq:marginal_gain}), we can obtain
 $I(S\cup\{u\})-I(S)\geq I(T\cup\{u\})-I(T)$ if $S
\subseteq T$. Therefore,  $I(S)$ is a submodular function. \done

\bibliographystyle{abbrv}
\bibliography{ref}

\begin{thebibliography}{10}

\bibitem{yelpdataset}
{Yelp Dataset}.
\newblock \url{https://www.yelp.com/dataset_challenge/dataset}.

\bibitem{barbieri2012topic}
N.~Barbieri, F.~Bonchi, and G.~Manco.
\newblock Topic-aware social influence propagation models.
\newblock In {\em Proc. of ICDM}, 2012.

\bibitem{Borgs2014Maximizing}
C.~Borgs, M.~Brautbar, J.~Chayes, and B.~Lucier.
\newblock {Maximizing Social Influence in Nearly Optimal Time}.
\newblock In {\em Proc. SODA}, 2014.

\bibitem{chen2010scalable}
W.~Chen, C.~Wang, and Y.~Wang.
\newblock {Scalable Influence Maximization for Prevalent Viral Marketing in
  Large-Scale Social Networks}.
\newblock In {\em Proc. ACM KDD}, 2010.

\bibitem{chen2009efficient}
W.~Chen, Y.~Wang, and S.~Yang.
\newblock {Efficient Influence Maximization in Social Networks}.
\newblock In {\em Proc. of ACM KDD}, 2009.

\bibitem{chen2010scalableICDM}
W.~Chen, Y.~Yuan, and L.~Zhang.
\newblock {Scalable Influence Maximization in Social Networks Under The Linear
  Threshold Model}.
\newblock In {\em Proc. of IEEE ICDM}, 2010.

\bibitem{everett1999centrality}
M.~G. Everett and S.~P. Borgatti.
\newblock {The Centrality of Groups and Classes}.
\newblock {\em The Journal of Math. Soc.}, 23(3):181--201, 1999.

\bibitem{freeman1979centrality}
L.~C. Freeman.
\newblock {Centrality in Social Networks Conceptual Clarification}.
\newblock {\em Social Networks}, 1(3):215--239, 1979.

\bibitem{galhotra2016holistic}
S.~Galhotra, A.~Arora, and S.~Roy.
\newblock Holistic influence maximization: Combining scalability and efficiency
  with opinion-aware models.
\newblock {\em arXiv preprint arXiv:1602.03110}, 2016.

\bibitem{hoeffding1963probability}
W.~Hoeffding.
\newblock {Probability Inequalities for Sums of Bounded Random Variables}.
\newblock {\em Journal of The American Statistical Association},
  58(301):13--30, 1963.

\bibitem{Jamali10_flixster}
M.~Jamali and M.~Ester.
\newblock {A Matrix Factorization Technique with Trust Propagation for
  Recommendation in Social Networks}.
\newblock In {\em Proc. of ACM RecSys}, 2010.

\bibitem{kempe2003maximizing}
D.~Kempe, J.~Kleinberg, and {\'E}.~Tardos.
\newblock {Maximizing The Spread of Influence Through a Social Network}.
\newblock In {\em Proc. of ACM KDD}, 2003.

\bibitem{kusumoto2014scalable}
M.~Kusumoto, T.~Maehara, and K.-i. Kawarabayashi.
\newblock {Scalable Similarity Search for Simrank}.
\newblock In {\em Proc. of ACM SIGMOD}, 2014.

\bibitem{lin2015analyzing}
Y.~Lin and J.~C. Lui.
\newblock {Analyzing Competitive Influence Maximization Problems With Partial
  Information: An Approximation Algorithmic Framework}.
\newblock {\em Performance Evaluation}, 91:187--204, 2015.

\bibitem{page1999pagerank}
L.~Page, S.~Brin, R.~Motwani, and T.~Winograd.
\newblock {The PageRank Citation Ranking: Bringing Order to The Web}.
\newblock Tech. report, 1999.

\bibitem{Tang-etal12b}
J.~Tang, H.~Gao, and H.~Liu.
\newblock {m{T}rust: {D}iscerning Multi-Faceted Trust in a Connected World}.
\newblock In {\em Proc. of ACM WSDM}, 2012.

\bibitem{tang2015influence}
Y.~Tang, Y.~Shi, and X.~Xiao.
\newblock {Influence Maximization in Near-Linear Time: A Martingale Approach}.
\newblock In {\em Proc. of ACM SIGMOD}, 2015.

\bibitem{tang2014influence}
Y.~Tang, X.~Xiao, and Y.~Shi.
\newblock {Influence Maximization: Near-Optimal Time Complexity Meets Practical
  Efficiency}.
\newblock In {\em Proc. of ACM SIGMOD}, 2014.

\bibitem{zhao2014measuring}
J.~Zhao, J.~Lui, D.~Towsley, and X.~Guan.
\newblock {Measuring and Maximizing Group Closeness Centrality over
  Disk-Resident Graphs}.
\newblock In {\em Proc. of SIMPLEX}, 2014.

\bibitem{zhao2015tale}
J.~Zhao, J.~Lui, D.~Towsley, P.~Wang, and X.~Guan.
\newblock {A Tale of Three Graphs: Sampling Design on Hybrid Social-Affiliation
  Networks}.
\newblock In {\em Proc. of IEEE ICDE}, 2015.

\end{thebibliography}

\end{document}